\documentstyle[12pt]{article}
\begin{document}
\begin{center}
{\Large \bf $A_0$ condensate in QCD}\\
\vspace{1cm}
{\large O.A.~Borisenko,
\footnote{email: gezin@gluk.apc.org}}\\
{\large \it
N.N.Bogolyubov Institute for Theoretical Physics, Kiev 252143,
Ukraine}\\
\vspace{1cm}
{\large J.~Boh\'a\v cik,
\footnote{email: bohacik@savba.savba.sk}}\\
{\large \it
Institute of Physics Slovak Academy of Sciences, 842 28 Bratislava, Slovakia}\\
\vspace{1cm}
{\large V.V.~Skalozub
\footnote{email: skalozub@dsuni.dnipron.glas.apc.org}}\\
{\large \it
Dnipropetrovsk State University, Dnipropetrovsk 320625, Ukraine}
\end{center}
\vspace{.5cm}

\begin{abstract}
A survey devoted to $A_0$-condensate in gauge theories at high
temperature is presented. Both the theoretical foundations of the
spontaneously generated condensate and known methods of its
calculation are discussed. As most important consequence the
$SU(N)$  global symmetry breakdown is investigated in details.
Influence of $A_0$ on matter fields is studied in different
aspects. Some new results concerning this subject are reported as well.
\end{abstract}

\newpage

\section{Introduction and Motivation}
In the last few years an interest to problems of high-temperature QCD
has considerably increased. Among commonly discussed topics on
the deconfining phase transition and the chiral symmetry restoration,
new possible phenomena - generation of the gauge classical field
(the so-called $A_0$-condensate) and spontaneous breaking of the global
gauge symmetry caused by $<A_0>$ - have become of great importance.
A number of new essential results obtained in various approaches
as well as interesting ideas and hopes connected with the non-zero
vacuum value $<A_0> \neq 0$ impelled us to undertake an attempt
to summarize here the most essential achievements and unsolved questions
in this area of gauge theories.  To be more transparent, we are going to
formulate a general point of view both on physical nature and
mathematical aspects of these phenomena representing them in different
methods of calculations. We would like to discuss as well the most
exciting problems and possible ways of their solving. Naturally,
we are aware  that
it is a sufficiently difficult task to present a general survey concerning
a particular question in a so rapidly developing area as
gauge theories at finite temperature. Moreover,
it cannot be excluded that by the moment of appearing
our paper in the journal some questions discussed here will be solved.
The situation gets complicated by the fact that as far as we
know there is no (mathematically) strict proof at the present time
that $A_{0}$-condensate must fall at high temperature.
Nevertheless, we believe that calculation of $<A_0>$ by different
methods and, on the other hand, derivation of the most significant
consequences of such a condensate on multi-particle systems
make our attempt quite justified.
Besides, the appearance of $A_{0}$-condensate
and the breakdown of global gauge symmetry can undoubtedly lead
to significant improvement of our conception both of the high temperature
behaviour of the strongly interacting matter and of the physics
of the gauge theories on the whole and surely have
a connection to other problems currently under investigation
(as, for instance, infrared problem,  behaviour of
quarks at non-zero baryonic number, etc.). All these questions
will be considered in the paper.

Let us begin with a comprehensive consideration of some known facts
obtained from the studies of QCD. The Hamiltonian can be formally
written as the sum of the chromoelectric and the chromomagnetic terms
and has the following form in lattice version of the theory
\begin{equation}
H = \frac{g^{2}}{2a}E^{2} + \frac{1}{2ag^{2}}UUU^{+}U^{+} =
H_{E} + H_{B}.
\label{1}
\end{equation}
\noindent
At finite temperature the behaviour of chromoelectric fields
has been well studied both in the perturbative (in $g^{2}$) and
especially in the non-perturbative regions. As is generally known,
in the strong coupling approximation the main contribution to the partition
function results from the chromoelectric part in eq.(\ref{1}), because
the chromomagnetic term being proportional to $g^{-2}$ can be treated
perturbatively in $\frac{1}{g^{2}}$. At high temperature because of periodic
boundary conditions the gauge field configurations called Polyakov loops
develop a non-vanishing expectation value, which breaks $Z(N)_{gl}$ symmetry
of the initial QCD-action and leads to the deconfinement. As the Polyakov loops
transform non-trivially under $Z(N)_{gl}$ rotations, their
non-zero expectation value could mean screening of $Z(N)$-charges
(or static quarks) at $T > T_{c}^{D}$ (where $T_{c}^{D}$  is the critical
temperature of the deconfinement phase transition). If this is the case,
one may claim that there exists a physical quantity which
characterizes the phenomenon of screening. This quantity is
called the Debye mass and is defined in the continuum as the zero
momentum limit of the zero-zero component of the vacuum polarization
tensor,
\begin{equation}
m^{2}_{D}(T) = - \Pi_{00}(\vec{k} \rightarrow 0, k_{0} = 0).
\label{2}
\end{equation}
\noindent
This definition gives a gauge invariant value only in the lowest non-trivial
order of the weak-coupling expansion for $SU(N_{c})$ gauge theory
with $N_{f}$ massless fermions
\begin{equation}
- \Pi_{00} = (\frac{N_{c}}{3} + \frac{N_{f}}{6})g^{2}(T) + O(g^{3}).
\label{3}
\end{equation}
\noindent
Calculation of Eq.(\ref{2}) even in the two-loop approximation
reveals infrared divergences and gauge dependence \cite{1nad}
leading to the conclusion that such a calculation cannot be
trusted as high-order corrections are not calculable systematically.
The starting point of the lattice studies is the connected
correlation function
\begin{equation}
\Gamma (R,T) = <W(0)W(R)> - <W(0)>^{2}
\label{4}
\end{equation}
\noindent
of the two Polyakov loops
\begin{equation}
W(\vec{x}) = Sp\prod_{t=1}^{N_{t}} U_{0}(\vec{x},t)
\label{5}
\end{equation}
\noindent
separated by a spatial distance $\vec{R}$. The colour averaged
quark-antiquark potential $V(R)$ defined by this correlation
\begin{equation}
V(R) = - \ln \Gamma (R,T)
\label{6}
\end{equation}
\noindent
has been found to be of the screening form \cite{2}
\begin{equation}
V(R) \approx \frac{const}{R} \exp(-m_{D}R)
\label{7}
\end{equation}
\noindent
In such a way we obtain an other definition of the Debye mass.
In the weak coupling approximation the definitions (\ref{2}) and (\ref{7})
have to coincide \cite{3nad2}.

Results of lower accuracy have been obtained from studies of
the chromomagnetic fields behaviour at high temperature. First of all,
two essential problems being important for understanding
the finite-temperature physics of non-abelian fields on the whole
are to be emphasized: infrared problem and area law for spatial Wilson loop.
The infrared problem appears
at attempts to study perturbatively the physics of the high
temperature region: The static chromomagnetic sector
develops infrared divergences which make
the perturbative expansion invalid starting from $g^{6}$ order. Leading
infrared divergences are those of the three-dimensional Yang-Mills theory
\cite{4lind}. It is widely believed that some physical characteristic
should exist leading to screening of the chromomagnetic forces and to
curing of the divergences. It is usually called the chromomagnetic mass
though there are several definitions of this quantity in the literature.

1) By analogy with the definition of the Debye mass (\ref{2})
the same limit of the space components of the $\Pi_{\mu \nu}(k_{0},\vec{k})$
could be considered:
\begin{equation}
m^{2}_{\mu}(T) = - \Pi_{ii}(\vec{k} \rightarrow 0, k_{0} = 0)
\label{8}
\end{equation}
\noindent
However, this limit was found to be equal to zero in the one-loop
approximation \cite{5kal}. Two-loop calculations performed in the axial gauge
have given finite result (but different in various schemes)
\cite{6kap}. Besides, specific difficulties of the axial gauge
(coming from its singular character) bring an ambiguity in obtained results
and, what is more important, lead to qualitatively different behaviour
of $\Pi_{ii}$ in comparison with relativistic gauges. The most
general two-loop calculations of $\Pi_{ii}$ in an arbitrary relativistic
gauge were done in \cite{7skl1}
\begin{equation}
- \Pi_{ii}^{(2)}(\vec{k},k_{0} = 0) = \frac{\pi g^{4}N_{c}^{2}}
{8(2\pi )^{3}\beta^{2}}[\frac{\alpha^{3}}{3} + \frac{4\alpha^{2}}{3}
+ \frac{13\alpha}{3} - 2] \ln \frac{\vec{k}^{2}}{\mu^{2}}
\label{9}
\end{equation}
\noindent
where $\alpha$ is the gauge fixing parameter ($\alpha = 1$ corresponds to the
Feynman gauge). There is the infrared singularity in the last equation
when $\vec{k}^{2} \rightarrow 0$. One can see as well that the sign
of the right-hand side of (\ref{9}) can be changed by varying
$\alpha$. Hence one must conclude,  expression (\ref{9}) cannot
define a physical mass. A detailed
discussion of such a situation may be found in \cite{7skl1}.
Consequently, at the present time we do not possess any regular methods for
calculation of the chromomagnetic mass (defined in the sketched way).
That is why other parameters should also be discussed.

2) By analogy with the definition of the Debye mass (\ref{4})-(\ref{7})
the chromomagnetic mass can be determined from the corresponding correlation
functions. In this case it has to be the spacelike Wilson loop in the
adjoint representation. If $m_{\mu} \neq 0$, "heavy" gluon current
sources have to be screened. As far as we know such a calculation
has not been performed yet.

3) If gluons acquire a magnetic mass, the magnetic fields of test
charges will be screened. It means we can measure a monopole-antimonopole
potential with the charges in the center of the gauge group. Corresponding
Monte-Carlo simulations, although on the small lattices, were
presented in \cite{8De}. The "magnetic" potential really is of the screening
form and the chromomagnetic mass determined from this potential is in a good
agreement with the formula $m_{\mu} = Bg^{2}(T)T$, where $g^{2}(T)$ is
the coupling constant renormalized at the momentum scale $T$. The computations
were performed both for the Wilson pure $SU(2)$ gluodynamics and for the
Mack-Petkova modified $SU(2)$-model which does not contain dynamical
monopoles.

It seems to be a logical conclusion that all of these three definitions
determine the same physical quantity though we do not know
a strict proof of this statement.

Second important (and strict) result which should be understood and
obviously concerns the chromomagnetic fields was obtained by Borgs
\cite{9bor} (see also \cite{10pol}). Its gist is the following:
The spacelike Wilson loop obeys the area law at arbitrary high temperature.
In principle, this result does not look very strange. The chromomagnetic
sector in Eq.(\ref{1}) at finite $T$ does not generate (unlike the
chromoelectric part) any configuration transforming non-trivially
under $Z(N)_{gl}$-rotations. Therefore it seems natural that there
is no screening effect of the corresponding sources (currents)
being in the fundamental representation in the vacuum.
Two points we would like to emphasize here are the following:

1) area law explains the impossibility to apply the standard weak
coupling expansion at least to all modes of the chromomagnetic
sector because the area law can never be achieved in this expansion.

2) as is known from the duality relations \cite{11hoft},
\cite{12seil}, the area law for the Wilson loop implies the perimeter law
for the 't Hooft disorder parameter. In its line this has to signify
a screening potential for a monopole-antimonopole pair.

Thus we can see that despite the fact that there is no strict proof that
the chromomagnetic mass should exist we have to take into account
that some kind of screening of the chromomagnetic forces does exist.
We consider that the result by Borgs and the duality
relations surely lead to this conclusion.

The crucial question coming from these facts concerns the
nature of this screening. Undoubtedly, a mechanism of
such a screening implies the existence of the magnetic mass
in one of the meanings above.

Two possible mechanisms supplying gauge fields with mass
and leading to the screening of chromomagnetic forces
are available and discussed now in the literature: dielectric
structure of the QCD vacuum and the spontaneous breaking of
global gauge symmetry in the background $<A_{0}>$.

1) Dielectric vacuum.

The attempts to formulate a confinement model in QCD framework led
in the second half of the seventies to the so-called dielectric theory.
In this model the gauge action is accompanied by additional terms
corresponding to the auxiliary field $\rho$. If we calculate
Gauss' law from the action of this kind we shall formally find the
relation with dielectric field as in the electrodynamics of the
dielectrics.
The definition of the pure gauge model is the following \cite{mack3}
\begin{equation}
Z = \int \prod_{x,\mu}D(\Phi)\exp (-S(\Phi))
\label{10}
\end{equation}
\noindent
where
\begin{eqnarray}
S(\Phi) = \sum_{x}\sum_{\mu,\nu}\lambda_{\mu \nu} F_{\mu \nu}
F_{\mu \nu}^{\star} + \frac{m^{2}}{2}\sum_{x}\Phi_{\mu}(x)
\Phi_{\mu}^{+}(x) + \sum_{x}V[\rho(x)]
\label{11}
\end{eqnarray}
\noindent
\begin{equation}
\Phi_{\mu} = \rho_{\mu}U_{\mu}, U_{\mu} \in SU(N),
0 \leq \rho < \infty
\label{12}
\end{equation}
\noindent
\begin{equation}
D(\Phi ) = \rho^{3}d\rho D\mu (U),
\label{13}
\end{equation}
\noindent
$D\mu (U)$ is the invariant group measure. The dielectric field is introduced
via the auxiliary field $\rho$. $V(\rho)$ is a gauge invariant
potential for this field.  $\rho$ is a scalar
field in the colour space for $SU(2)$ and a tensor field for $SU(N > 2)$.
One of the most essential features of such theories
is a possibility to introduce a gauge-invariant mass
for gauge fields $\Phi_{\mu}(x)$ (see the second term in (\ref{11})).

It has been recently shown that the lattice effective
action for the infrared dangerous static modes appears
at the dimensional reduction to be a little complicated version of
the described dielectric theory \cite{ildgt}.
Physical mass of the dielectric field is
\begin{equation}
m_{d} = T(\lambda^{0} - 2 \lambda_{e} \cos^{2}(a_{\beta}g <A_{0}>))
\label{14}
\end{equation}
\noindent
where $\lambda^{0} = 2( \mbox{flat function of} \ T,g)$,
$\lambda_{e} = \frac{a_{\sigma}}{g^{2}a_{\beta}}$.
Undoubtedly, this mass leads to the screening of chromomagnetic forces
and therefore of all gluon sources being in the adjoint representation
\cite{ildgt}. It means in a sense the solution of the infrared problem,
truly,
perturbative expansion is invalid in any case since the dielectric field
appears only at the non-perturbative level and vanishes if we try to apply
the perturbative expansion.

2) $A_0$-condensate.

The second possibility is the spontaneous breaking of the
$SU(N)$ global symmetry at high temperature accompanied by the
generation of the chromomagnetic mass.
Just this phenomenon will be in the focus of our interest here.
The only known mechanism providing such a breakdown at high temperature is the
generation of $A_0$-condensate	which is a constant part of the
temporal gauge field component $A_{0} = const$ \cite{gross}.
The nature of this parameter is connected with special properties
of non-abelian gauge fields at finite temperature which display
themselves in different ways depending on the particular method
of calculations. We adduce below
a comprehensive description of these properties.
In general, the possibility of $<A_{0}> \neq 0$ originates
from the compactification of the imaginary time direction at $T \neq 0$.
So, it is necessary to combine periodicity in the imaginary time
$A_{\mu}(0) = A_{\mu}(\beta)$, $A_{\mu}^{\prime}(0) = A_{\mu}^{\prime}(\beta)$
and the gauge transformations of the fields
${A}_{\mu}^{\prime} = UA_{\mu}U^{+} + \frac{i}{g}U \partial_{\mu}U^{+}$ where
$U$ is the gauge transformation operator. To make this transformation agree
with the periodicity, the operator $U = \exp (igA_{0}\beta )$ (where $A_{0} =
A_{0}^{a}t^{a}$) must commute with generators $t^{a}$. Hence,
$A_{0}$ should belong to the center of the gauge group: $A_{0} =
\frac{2\pi n}{\beta g N}$, $n \in Z_{N}, n = 0,1,...,N-1$. Actually,
the vacuum value of $A_{0}$ has to be calculated from a full
effective action with quantum fluctuations included and if it will
be found that $<A_{0}>$ differs from $\frac{2\pi n}{\beta g N}$,
the spontaneous breaking of the  gauge	symmetry
will be determined. In what follows, speaking about $A_0$-condensate
we will assume it as a non-zero vacuum value of zero gauge field component
obtained at a minimum position of the full effective action.
The presence of the classical gauge field
in vacuum may give a necessary missing parameter for solving
the infrared problem. Moreover, this condensate effects various
processes at high temperatures. In particular, it may lead to the
spontaneous breaking of the baryon charge symmetry, etc.

As a matter of fact, the two screening mechanisms could be somehow connected.
So it has been discussed in \cite{ildgt} that the mass of the dielectric
field in the reduced theory can be proportional to $<A_{0}>$.
We shall consider this possibility in the corresponding part of the survey.

The described general idea can get different realization depending
on applied calculation methods. A general picture of the effects of the
$A_{0}$-condensate will appear when the results obtained in various
particular approaches will be gathered together. The goal of the present
review is to describe in a systematic way the results obtained
in three approaches to $A_{0}$-condensate calculations. We discuss
the method of effective Lagrangians for $A_{0}$-field,
the Hamiltonian description of the $SU(N)$-symmetry breaking in the
background $A_{0}$ both on the lattice and in the continuum theory,
and the loop expansion of the effective action in QCD.
In all these approaches the gauge field condensate
has been determined in the framework of the appropriate approximation
schemes. Anyway, at the present time there is no common opinion about
realization of this phenomenon in the nature because of a number
of problems concerning the accuracy, gauge invariance and precision of
the calculations. So, in what follows we are going to adduce the
comprehensive analysis of these problems and compare the results obtained.

Our Survey is organized as follows.

First we remind in brief the most essential features of
gauge theories at finite temperature and introduce our notations
(chapter 2).
The general status of $A_{0}$-condensate is presented in chapter 3.
In chapter 4 we go through the calculation of the condensate in the
Hamiltonian formulation on the lattice.
The gauge independence of the condensate in this approach is discussed
as well.
In chapter 5 the loop expansion for $<A_{0}>$ calculation is
examined. The central point of this examination is the proof
of gauge independence of the condensate by means of the
Nielsen identities.
Chapter 6 is devoted to elaborating of the effective Lagrangian
method for the expectation value of $A_{0}$.
In chapter 7 we compute the condensate in the theory with
dynamical quarks. We consider both lattice and loop approaches
and compare their results.
Some consequences of the non-zero $<A_{0}>$ are reviewed in
chapter 8. Here we are going to give some new results related to
the $A_0$-condensate phenomenon.
So we give a sketch of the calculations of the heavy
quark potential in the background $<A_{0}>$. The infrared problem
is reexamined at $<A_{0}> \neq 0$ in reduced lattice theory at high
temperature. We give a proof that the adjoint spatial Wilson loop obeys
perimeter law, which means the screening of chromomagnetic forces.
Further, we discuss the spontaneous breaking of the baryon charge symmetry
and some closely related problems. The general conditions for appearing
of the Chern-Simons action in the background $<A_0>$
at high temperature are presented as well.
Brief Summary and Discussion can be found in chapter 9.

\section{Gauge theories at finite temperature. General outlook}

In order to make our survey as self-contained as possible and for
convenience of readers we would like to make a sketch of $SU(N)$
gauge theories at $T \neq 0$. To begin with, we remind the general
formulation of the finite temperature theories both in the continuum
and on the lattice and introduce our notations.
In the continuum the QCD action has the form:
\begin{eqnarray}
S &=& \int [d^{3}x] \int_{0}^{\beta} dt L_{QCD},   \nonumber  \\
L_{QCD}&=&\frac{1}{2g^2}\ Tr{F_{\mu \nu} F^{\mu \nu}} +
\sum_{f=1}^{N_f} \overline{\Psi}^f [D_n\gamma_n-m_f+(D_0+i\mu)\gamma_0]\Psi^f,
\label{15}
\end{eqnarray}
\noindent
where
\begin{equation}
F_{\mu \nu}= \partial_{\mu}A_{\nu}\ -\ \partial_{\nu}A_{\mu} + ig
[A_{\mu},A_{\nu}],\ \ \ D_{\mu} = \partial_{\mu}-ig A_{\mu},\ \
A_{\mu}=A_{\mu}^a t^a,
\label{16}
\end{equation}
\noindent
$\mu$ is the baryonic chemical potential.

On the lattice we consider the Wilson action:
\begin{eqnarray}
S&=&S^G+S^f \equiv \nonumber \\
& & \frac{2N_c}{g^2} \sum_{P} [1-\frac{1}{2N_c}\ TrU(\partial P)]
+ C.C. + \nonumber \\
& &\sum_{x,n=-d}^{d}\sum_f \overline{\Psi}^f_x\ \Gamma_{x,x+n}U_n (x)
\Psi^f_{x+n}
+ \sum_{x} m_f\ \overline{\Psi}^f_x\ \Gamma \Psi^f_x
\label{17}
\end{eqnarray}
\noindent
where $U_n(x)\in SU(N)$ and $Tr\ U(\partial P)$ is the plaquette character
in the fundamental representation.
The form of the matrices $\Gamma_{x,x+n}$ and $\Gamma$ depends on a sort
of lattice fermions.

It is well-known that the problem of quantization of gauge theories
depends essentially on the boundary conditions imposed on the gauge
fields. The compact topology leads to another physical picture of the
gauge model than the Euclidean topology accompanied by the corresponding
boundary conditions.
QCD at zero temperature is studied in the Euclidean space $G=R^d$.
A transition to the cylinder topology is achieved by imposing the periodic
boundary conditions on the gauge fields along the imaginary time direction
with period $\beta =1/T$ which plays the role of inverse temperature.

Thus, the finite temperature theories are defined on the space
$G=R^{d-1}\otimes S^1$.
The periodic boundary conditions generate the new observable
known as the Polyakov loop \cite{polonyi}:
\begin{equation}
W = \ P\ \exp\ (ig\int_0^{\beta}dt\ A_0(x,t))
\label{18}
\end{equation}
\noindent
(or see (\ref{5}) for the lattice definition).

The finite temperature formalism is used for studying  the
thermodynamical features of QCD and its different  phases. So,
let us discuss in brief these features and, first of all,
the phase structure obtained from lattice calculations.

Nowadays two  phase  transitions  are usually considered to be
relevant to QCD at  finite  temperature
and/or	baryon density. One is the deconfinement phase
transition, i.e. transition  from  the	confining  phase  of  the
hadronic  matter to the quark-gluon plasma phase.  The other is
the chiral symmetry restoration  phase	transition.
As is generally believed,  the physical
picture  obtained  from lattice  studies  could   be   the
following. The pure gluonic QCD action has the global $Z(N)$ symmetry,
which leads to the confinement of color quark  charges	in  the  low
temperature  phase. In the high temperature phase $Z(N)$ symmetry
is spontaneously broken and quarks are screened by some specific
configurations	of  the gluon fields.  This transition is called
the deconfinement phase transition and it is best  studied in pure
gluodynamics.
Yaffe and Svetitsky \cite{yasve} have proved the relation between
the free energy of the infinitely heavy quark $F_{q}$ and the expectation
value of the trace of the Polyakov loop, $<TrW> = \exp (-\beta F_{q})$.
This relation and the behaviour of $TrW_x$ under
$Z(N)_{gl}$ transformations
\begin{equation}
TrW_{x} \longrightarrow zTrW_{x}
\label{19}
\end{equation}
were used as the basic ones to study  the deconfinement phase transition
in the Yang-Mills theory\cite{mcler}. Difference  of the expectation
value of the $TrW_{x}$ from zero is the signal of the spontaneous
breaking of $Z(N)_{gl}$ symmetry and also of the deconfining
of the static colour charges transforming non-trivially under $Z(N)$:
\begin{eqnarray}
\langle \{N^{-1}TrW_{x}\}\rangle = \left\{  \begin{array}{c}
	       0, T<T_{c}^{D},\ confinement\ phase,  \\
		  Z\ast f(T),\ T>T_{c}^{D},\ Z\in Z(N)_{gl},\ f(T)\leq 1,
				       \ deconf.\ phase \end{array} \right.
\label{20}
\end{eqnarray}
In the full QCD with dynamical quarks  $Z(N)$  global  symmetry  is
evidently  broken  from  the beginning	and  no  other	appropriate
order parameters for this transition are known.
On the other hand, in the  presence  of the
dynamical fermions the first order deconfining transition in the
pure $SU(3)$ gauge  theory  is	getting  weaker  as the quark  mass
decreases. This  weakening is going on up to the quark mass of
order $T$.  However,  for smaller masses the transition  has  been
found  to  become  stronger  with  the quark  mass decreasing.	This
transition can be characterized  by  the  chiral order parameter
$< \sigma >$ and is called  the  chiral phase transition.  Numerical
results show that the deconfinement and the restoration  of the chiral
symmetry, being distinctive by character at the first sight, are
in fact indistinguishable in the region  of intermediate  quark
masses.

Before proceeding further to discuss the gluon field condensation
at high temperature let us make a short summary on the perturbative
QCD vacuum at $T \neq 0$ (see \cite{gross}, \cite{kisl}, \cite{mcler1}
for a detailed review of the finite temperature properties of QCD)
The perturbative
vacuum is supposed to have no structures or condensates.
Its main property is the asymptotic freedom ($g(T) \rightarrow 0$
in the limit $T \rightarrow \infty$ \cite{gross}).
At $T \neq 0$ the momentum space is naturally divided in
two parts: 1) $\mid k \mid \gg gT$ where the perturbative methods and
results  are to be reliable owing to asymptotic freedom;
2)$\mid k \mid \ll gT$ , this is a
truly infrared region. Just for these momenta one runs into
infrared divergencies in the higher orders of perturbation theory,
gauge dependence and other problems which signal that the
perturbative vacuum is not adequate to the nature. So, as is
expected and has been discussed in the introduction,
some new macroscopic parameter (like the Debye mass)
should be dynamically generated. As a candidate, the gluon magnetic
mass, calculated by perturbative and non-perturbative methods \cite{kalash5},
\cite{5kal} has been discussed.

At the same time another possibility has also originated from
the analysis by perturbative methods. Here we have in mind the dimension
reduction process at high temperature \cite{gross}, \cite{app1},
\cite{app2}, \cite{land}. The basic reason  is as follows:
in the gluon free propagator $\frac{1}{k^{2} + (2 \pi nT)^{2}}$ of
imaginary-time perturbation theory, the term $2 \pi nT$ acts like a
"mass" in the three-dimensional theory. According to
the Appelquist-Carazzone theorem \cite{app1} all nonstatic modes $n \neq 0$
decouple in the
limit $T \rightarrow \infty$, leaving the static ($=$ three-dimensional)
sector as the effective theory.
This idea can be realized in different ways and,
actually, a number of problems of calculation of the effective
Lagrangian appeared \cite{land}.  But in general it gives a possibility to
search for a non-trivial vacuum at finite temperature. In the present
paper the problem of the investigation of the gauge field vacuum via
the described mechanism will be one of the discussed topics
both in continuum QCD and beyond perturbative horizon on the lattice.

\section{General status of $ A_0 $-condensate}

In this chapter we would like to give general reasons for condensation of the
electrostatic potential and derive the corresponding mathematical
foundation. We formulate a mathematical task of calculation of
$<A_{0}>$ and give an outlook on the previous investigations.

Firstly, we consider the lattice gauge theory without periodic boundary
conditions (QCD at $T=0$) with the partition function of the form:
\begin{equation}
Z=\int D\mu (U) D\bar{\Psi} D\Psi \exp (-S^G -S^F).
\label{21}
\end{equation}
\noindent
The gluonic $S^G$ and the fermionic $S^F$ parts of the action are
expressed in eq.(\ref{17}). The gauge field integration is performed
over $SU(N)$-invariant group measure $D\mu(U)$. Due to this fact the results
of the calculations are gauge independent. For example,
$$
\langle B \rangle = \langle B(U) \rangle_{U_0 = 1\ (A_0=0)}, \nonumber	\\
$$
\noindent
if $ B(U)$ is a gauge invariant function\cite{12seil}. This means that there
is a gauge transformation which allows us to fix gauge $U_0=1$ or
$A_0=0$. We can easily see that any constant $\langle A_0\rangle $ may be
singled out by an appropriate gauge transformation.

The partition function at finite temperatures is calculated
within the following periodic boundary conditions:
\begin{eqnarray}
U_\mu(x,\tau)&=&U_\mu(x,\tau+N_{\beta}),  \nonumber\\
\Psi(x,\tau)&=&-\Psi(x,\tau+N_{\beta}),  \\
\bar{\Psi}(x,\tau)&=&-\bar{\Psi}(x,\tau+N_{\beta}) \nonumber
\label{22}
\end{eqnarray}
\noindent
where $N_{\beta}$ is a number of the lattice sites in the time direction.
These boundary conditions are incompatible with the gauge $A_0=0$. We may fix
a static diagonal gauge only. Let $U_0(x,\tau)=V_x$, where $V_x$ is
a diagonal time-independent $SU(N)$ matrix. After performing the gauge
transformations
\begin{eqnarray}
U_n(x,\tau) \longrightarrow (V_x)^{\tau}U_n(x,\tau)(V_{x+n})^{-\tau}
\nonumber   \\
U_{n}^{+}(x,\tau) \longrightarrow (V_{x+n})^{\tau}U_{n}^{+}(x,\tau)(V_x)^
{-\tau}       \nonumber \\
 n=1,...,d    \nonumber \\
\Psi(x,\tau) \longrightarrow (V_x)^{\tau}\Psi(x,\tau)  \nonumber   \\
\bar{\Psi}(x,\tau) \longrightarrow \bar{\Psi}(x,\tau)(V_x)^{-\tau}
\label{23}
\end{eqnarray}
(which lead to the gauge $A_0=0$ at zero temperature) we can remove the
matrices $V(x)$ in the action from all links but the last one where
all of them are grouped gathering in the Polyakov loops $W_x$. On the
lattice in the static gauge the Polyakov loop becomes
\begin{equation}
W_x = P\prod_{\tau=1}^{N_\beta }U_0(x,\tau ) = \exp(i\beta gA_0(x)).
\label{24}
\end{equation}
In such a way we obtain the new terms in the chromoelectric and
in the fermionic parts of the action
\begin{eqnarray}
S^G(W) &=& \frac{2}{g^2}\sum_{x,n}Re\ Tr\ W_{x}U_{n}(N_{\beta}-1,x)\cdot
W_{x+n}^{+}U_{n}^{+}(1,x) , \nonumber	 \\
\label{25}
S^F(W) &=& \sum_{x;\pm 0}\bar{\Psi}_{x}(N_{\beta}-1)\Gamma_{x,x\pm 0}\cdot
W_{x}(\pm 0)\Psi_{x}(1) ,  \\
     W_{x}(\pm 0) &=& \left\{  \begin{array}{c} W_{x} \\
					      W_{x}^{+} .
			      \end{array} \right.	   \nonumber
\end {eqnarray}
\noindent
These terms are absent in the theory without periodic boundary conditions and
describe the interactions between gauge fields, fermionic fields and
the Polyakov loops. Owing just to these new terms the $SU(N)$ gauge
theory has the non-trivial phase structure described above. Moreover,
we can deduce from the last equations a possibility of the
spontaneous breaking of the $SU(N)$ global gauge symmetry caused by
condensation of the chromoelectric potential.
The symmetry group of the pure gauge theory is the direct product
of the group of the local gauge symmetry, the group of the
global gauge symmetry and its center subgroup:
$SU(N)_{loc}\times SU(N)_{gl}\times Z(N)_{gl}$.
The last one acts only on the Polyakov loops.
The fermionic part  of the action (\ref{25}) violates the
global center symmetry explicitly, the symmetry group becomes
$SU(N)_{loc}\times SU(N)_{gl}$ and the expectation value of the $TrW_x$
is not the appropriate order parameter at all\cite{banks}. Nevertheless
the eigenvalues of the Polyakov loop
\begin{equation}
		     W_{x}^{l} = \exp(i \varphi_{l}(x))
\label{26}
\end{equation}
\noindent
can be used for studying the behaviour of the system under
$SU(3)_{gl}$ symmetry because the Polyakov loop transforms under
$SU(3)$ rotations like the matter fields in the adjoint
representation\cite{polonyi}
\begin{equation}
	      W_{x} \longrightarrow U\ W_{x}\ U^{+},\ \ U\in SU(N)
\label{27}
\end{equation}
\noindent
The eigenvalues of the Polyakov loop can be chosen, for example, in the
$SU(3)$ theory as
\begin{eqnarray}
(\beta g)^{-1}\varphi_{1}&=& A_{0}^{3}+
\frac{1}{\sqrt{3}}A_{0}^{8},	 \nonumber \\
(\beta g)^{-1}\varphi_{2}&=& -A_{0}^{3}+
\frac{1}{\sqrt{3}}A_{0}^{8}, \nonumber \\
\sum_{l=3}^{N}\varphi_{l}&=& 0					 \nonumber
\label{28}
\end{eqnarray}
\noindent
Therefore the expectation value of $W_{x}^{l}$ may be different from zero
only if the spontaneous breaking of the $SU(3)_{gl}$ symmetry is possible.
The constant field $A_0$ can be cancelled in the action  only  when
$W(\langle A_{0}\rangle )$ belongs to $Z(N)$, as one may easily see from eq.
(\ref{25}) because in this case the matrices $W$ commute with $U_n$.
In any other case the constant $A_0$ will be present in the action
and the global $SU(3)$-symmetry is broken up to its Cartan subgroup
\begin{equation}
W(\langle A_{0}\rangle ) \ni Z(N) \Longrightarrow SU(N) \longrightarrow
[U(1)]^{N-1}
\label{29}
\end{equation}
\noindent
It is clear now that such a mechanism can work both in the pure gluodynamics
and in the theory with dynamical quarks unlike the mechanism of the
spontaneous breaking of the $Z(N)_{gl}$ symmetry.

Thus, we have two possibilities to quantize the gauge theory in the
space $G=R^{d-1}\otimes S^1$:

1) The theory in $G=R^{d-1}\otimes S^1$ must always possess the same
symmetry of the vacuum and the same Lagrangian as the theory in the Euclidean
space $G=R^{d}$. So, for instance, any constant $A_0$ should be
singled out by the corresponding gauge transformation from eq.(\ref{23}).
This constraint leads to the restriction $W_{x}^{l} = \exp (\frac{2\pi i}{N}
q_{l}(x)), q = 0,1,...,N-1$ and for the quantum theory we obtain
\begin{equation}
Z= \sum_{q_{l}(x)}\int_{U_{n}(\tau) = U_{n}(\tau + \beta)} D\mu (U)
D\bar{\Psi} D\Psi \exp (-S(\exp (\frac{2\pi i}{N}q_{l}(x)), U,
\bar{\Psi},\Psi)).
\label{30}
\end{equation}
\noindent

2) The global symmetries of the vacuum are determined by dynamics
of the gauge system itself. Then,  in eq.(\ref{30}) we have
$\sum_{q_{l}(x)} \rightarrow \int d\mu (W_{x})$, where $d\mu$ is the
invariant $SU(N)$ measure.

In the former case there is the only
possibility of the spontaneous breaking of the global $Z(N)$ symmetry
whereas in the latter one the spontaneous breaking of both $Z(N)$ and
$SU(N)$ global symmetry can occur. It is difficult to give a preference
to any of these formulations from the theoretical point of view.
The second formulation commonly seems to be more relevant.
In what follows just this formulation will be explored here.

Thus, from the picture formulated above we come to a task of calculating
$\langle A_0\rangle$ which can be formulated in the following way:
The configurations $A_n = 0\ (n=1,...,d),\ A_0 = const$ are the solutions
of the Yang-Mills equations with the additional condition that
the classical action equals
zero. At the classical level, however, the breakdown of the symmetry is
absent and the system is in the symmetry invariant minimum. Can the
quantum fluctuations generating the effective potential for $A_0$ lead
to spontaneous breaking of the global symmetry and produce the minima
of the effective action with a non-trivial value of $\langle \beta gA_0\rangle
\neq \frac{2q\pi}{N}$ ? On the lattice the situation is the same: for the
configurations $U_n = 1,\ W_l = const,$ the classical lattice action
$S^G$ is equal to zero and therefore we can ask the same question.

Actually, the investigation of the $A_{0}$-condensate started
ten years ago and $<A_{0}> \neq 0$ was determined in various approaches.
Nevertheless, at the moment we have neither a strict proof that
the condensate has to appear nor a common opinion about its generation.
This is mainly due to the mathematical difficulties which have been
encountered in the used approaches. For example, in the loop expansion
method $<A_{0}> \neq 0$ is derived from the two loop effective action
$W(A_{0}, \xi )$ \cite{belyev},\cite{enq},\cite{anish},\cite{belyev2},
\cite{belyev3},\cite{scl8},\cite{scl2},\cite{kalnew},\cite{chub},
\cite{sclnew}. So, to verify the result the three-loop contribution
should be evaluated. This very complicated task has not been solved yet.
Other approaches discussed in the literature \cite{bor},\cite{polvaz},
\cite{bogacek},\cite{olesz},\cite{pisar} also contain either some
uncertainties or  unsolved problems.

Recently, the possibility of $A_{0}$ condensation has been called in
question from the point of view of its gauge invariance\cite{belyev2}. As was
found in the background $R_{\xi}$ gauge, both the effective action and its
minimum value appear to be dependent on the gauge fixing parameter
$\xi$ \cite{enq},\cite{scl8},\cite{belyev3}. Hence, a doubt about gauge
invariance of the phenomenon has arisen. To resolve this doubt in the
perturbation theory several methods for gauge invariant calculations
have been conjectured\cite{belyev3},\cite{kalnew},\cite{bhat}.
But in all of them the result -  no  real condensation at
two loop level - has been stated. On the other hand, the gauge invariant
results of the analytical lattice investigations \cite{bor}
unambiguously show that condensate does appear. The same conclusion
has been supported via the Nielsen identities method \cite{niels} at
the two loop level (in order $g^{2}$ in coupling constant). Another
approach to the problem of gauge invariance of the condensate was
recently proposed in\cite{pisar} where the authors built a partition
function for the eigenvalues of the Polyakov loop in the Hamiltonian
formulation of the continuum theory. The result $A_{0} = 0$ was obtained.
We shall discuss this result later on remarking that this conclusion
is in obvious contradiction with apparently gauge invariant lattice
calculations \cite{polonyi},\cite{bor}.

The situation needs to be clarified in any way because $A_{0}$-condensate,
if it does realize in the nature at high temperature, would be a very
essential element of the self-consistent finite temperature gauge theory.
It seems to us that the only way to comprehend the situation
is to fix the strong results of various approaches
and to find out both their common point and sources of discrepancies.
At the same time it would be very desirable to find crucial phenomena
connected to $<A_{0}> \neq 0$ which can be detected in future experiments
on the heavy-ion collisions. These are main two purposes of the present
paper. We believe that being gathered together,
the results of various approaches can help to elucidate a lot
of difficulties and outline the prospects for future investigations.

\section{Hamiltonian formulation of the gauge theories and
	 $A_{0}$-condensate on the lattice}

The first approach we would like to analyze here is the Hamiltonian
formulation of the gauge theories. It is very desirable  to have
an apparently gauge invariant approach to defining and calculating
$<A_{0}>$-condensate. Possessing this essential property,
the Hamiltonian approach allows us to be convinced that $A_{0}$
condensation will be a gauge invariant phenomenon. On the other hand,
we want to have a strict proof that $<A_{0}> \neq 0$ in the
framework of the reliable approximation scheme. Fortunately,
such an approximation does exist. It is the strong coupling
expansion in the lattice theories. In what follows we
consider, at first, the strong coupling region of the Hamiltonian lattice
formulation. Then, the continuum version of the Hamiltonian
approach will be discussed.

The detailed description of the lattice Hamiltonian formulation
can be found in the series of the papers \cite{suss}, \cite{polyakov}
and in the review \cite{kogut}. The Hamiltonian approach on the lattice
was developed for the first time in \cite{kogsus}.
In the construction of the Hamiltonian
partition function we follow our own method \cite{borhap} adducing a proof
of its equivalent to earlier methods. The Hamiltonian of the lattice
gluodynamics  in the strong coupling approximation includes only
the chromoelectric part
\begin{equation}
H = \sum_{links} (\frac{g^{2}}{2a})E^{2}(l)
\label{31}
\end{equation}
\noindent
where $E(l) = i\partial / \partial (A_{l})$ - are the chromoelectric field
operators. In this approach the chromomagnetic term can be treated
perturbatively at $g^{2} \rightarrow \infty $.
Calculation of the partition function
\begin{equation}
Z = \tilde{Sp} \exp (-\beta H)
\label{32}
\end{equation}
\noindent
is connected with summing over local gauge-invariant
states. This is reflected in sign $\tilde{Sp}$ in (\ref{32}).
The corresponding physical Hilbert space is determined by
Gauss' law. To satisfy the latter the usual method is to introduce
the necessary $\delta$-function. In order to solve this task we use the more
general method connected with the projection operator technique. It should be
emphasized at once that the Hamiltonian formulation allows to obtain
an effective action for the eigenvalues of the Polyakov loop in the gauge
$A_{0} = 0$. In what follows we are going to demonstrate that carrying out
the projection onto local gauge invariant states and summing over
the eigenvalues of the diagonal operators of the gauge group are
equivalent to integration over spatial gauge fields in the Euclidean version
of the lattice theory without fixing any gauge. It is a rather important
point which was missed, for instance, in Ref. \cite{pisar}.

Our starting point is the partition function (\ref{32}) where we implant
a projection operator in each lattice site
\begin{equation}
Z = Sp{ \exp (-\beta \sum_{links} (\frac{g^{2}}{2a})E^{2}(l)) P},
P = \prod_{x}P_{x}^{0}
\label{33}
\end{equation}
\noindent
where the operator $P^{r}$ is defined as
\begin{equation}
P^{r} = \frac{h_{r}}{\mid g \mid} \sum_{j=1}^{\mid g \mid}
\Omega_{j}^{\star ,r}(g) U_{j}(g).
\label{34}
\end{equation}
\noindent
To make the next mathematical construction more transparent we have put down
here $P^{r}$ for a discrete group, $h_{r}$ is the dimension of the
representation
$r$, $\mid g \mid$ is the rang of the group and
$\Omega^{r}$ is the character of the irreducible representation $r$.
The matrices $U_{j}(g)$ are located on the
links. In eq.(\ref{34}) we must consider all representations which can be
combined in such a way to form singlets in each lattice site. Eq. (\ref{34})
can be rewritten to the form
\begin{equation}
P^{r} = \frac{h_{r}}{\mid g \mid} \sum_{k=1}^{n} \Omega_{k}^{\star ,r}
\sum_{j=1}^{g_{k}} U_{k,j} =
\frac{h_{r}}{\mid g \mid} \sum_{k=1}^{n} \Omega_{k}^{\star ,r} C(k).
\label{35}
\end{equation}
\noindent
The second identity is the definition of the class operator $C(k)$ which
contains $g_{k}$ elements. $n$ is the number of linearly independent
class operators of the representation group $G$ \cite{grouprep}.
There exists a representation for $C(k)$
\begin{equation}
C(k) = \frac{g_{k}}{\mid G \mid} \sum_{j=1}^{\mid G \mid} U_{j}U_{k}U_{j}^{-1}
\label{36}
\end{equation}
\noindent
from which it is easy to deduce the most important
property of the class operator: it is invariant under transformations
out of group of representation $G$ or in other words
\begin{equation}
    [ C(k),U(g) ] = 0.
\label{37}
\end{equation}
\noindent
A generalization of (\ref{35}) and (\ref{36}) on the Lie group is not
very complex and is founded on using the so-called "unitary Weyl's trick".
As is known, any unitary representation of the Lie group may be presented
in the form \cite{weyl}
\begin{equation}
U(\Phi) = V(v) \xi (\phi) V^{-1}(v)
\label{38}
\end{equation}
\noindent
by the appropriate choice of the unitary coordinate system. In these terms
the invariant measure is
\begin{equation}
d \Phi = dvd \mu (\phi),
\label{39}
\end{equation}
\noindent
which allows to integrate over $\nu$-variables.
After that we have for $P^{r}$
\begin{equation}
P^{r} = h_{r} \int d \Phi \Omega^{\star ,r}(\Phi) U(\Phi) =
h_{r} \int d \mu (\phi) \Omega^{\star ,r} \int d \nu U(\phi ,\nu).
\label{40}
\end{equation}
\noindent
For class operator we obtain in the case of the Lie group
\begin{equation}
C(\phi ) = \int d \nu U(\phi ,\nu).
\label{41}
\end{equation}
\noindent
It is obvious that $C(\phi)$ possesses the property of the invariance
analogous with (\ref{37}). Just this property is the most important
one in the selection of the local gauge-invariant states. Let $C_{l}$
be the eigenvalues of the Hamiltonian (\ref{31}) and $U_{\mu}^{l}(x)$
be the eigenfunctions of $E^{2}$. They are taken as variables  on
the lattice links, $x$ is the lattice site and $\mu$ is the unit vector.
Then, the partition function (\ref{33}) is rewritten in the form
\begin{equation}
Z = \sum_{l} K_{l} \exp (-\gamma C_{l}) ,
\label{42}
\end{equation}
\noindent
where $\gamma = \frac{\beta g^{2}}{2 a_{\sigma}}$ .
The operator $K_{l}$ selects gauge-invariant states among
all eigenstates of the Hamiltonian and has the form
\begin{equation}
K_{l} = \int d \mu (U) \prod_{x,\nu}U_{x,\nu}^{+,l} P
\prod_{x,\nu}U_{x,\nu}^{l}.
\label{43}
\end{equation}
\noindent
 From above formulae it follows that
\begin{equation}
P \prod_{x,\nu}U_{x,\nu}^{l} = \int \prod_{x}d \mu (\phi_{x}) \prod_{x,\nu}
\Omega (\phi_{x}) U_{x,\nu}^{l} \Omega^+ (\phi_{x+\nu}).
\label{44}
\end{equation}
\noindent
The operator $P$ satisfies the relation $PP = P$ and
is nothing but the projection operator.
How it is working may be seen from the last three equations. Next, we would
like to describe how to connect this technique with standard projection
and with Gauss' law.

Let us choose the transformations $\Omega(\phi_{x})$ in the form
\begin{eqnarray}
\Omega(\phi_{x})U_{x,\nu}&=&
\exp (iE_{x,\nu}^{a} \phi_{x}^{a}) U_{x,\nu},	\nonumber   \\
U_{x,\nu} \Omega^{+}(\phi_{x+\nu})&=&
U_{x,\nu} \exp (- iE_{x+\nu, \nu}^{a} \phi_{x}^{a}),
\label{45}
\end{eqnarray}
\noindent
where the generators $E_{x,\nu}^{a}$ are the chromoelectric field
tension operators defined on the lattice links. Then, grouping
$E_{x,\nu}^{a}$ at the same $\phi_{x}^{a}$ we obtain that
\begin{equation}
P = \prod_{x}P_{x}^{0}, \
P_{x}^{0} = \int d \mu (\phi_{x}) \exp (-iQ^{a}_{x} \phi_{x}^{a})
\label{46}
\end{equation}
\noindent
where
\begin{equation}
Q^{a}_{x} = \sum_{\nu} E^{a}_{x, \nu}
\label{47}
\end{equation}
\noindent
are the colour charge operators and the generators of the gauge transformations
in the lattice sites.  $P_{x}^{0}$ in Eq. (\ref{46})
is the projection operator
onto states with zero charge and is a particular choice of the operator
(\ref{40}).

The connection of Gauss' law $\delta (\Delta E)$ with a projection operator
technique is not a trivial task as the calculations are quite complicated
because the operators $Q^{a}$ do not commute with each other and,
consequently, it is impossible to find eigenfunctions for all $Q^{a}$
at the same time. Applying  "Weyl's trick" again we present
$\delta (Q^{a})$ in the form
$$
\delta (Q^{a}) = \int d^{(N^{2}-N)} \nu d^{(N-1)} \phi
\tilde{V}(\nu) \exp [i \sum_{a}^{N-1}\phi^{a}Q^{a}] \tilde{V}^{+}(\nu).
$$
In the last equation the matrices $\tilde{V}(\nu)$ diagonalize
$\exp [i \sum_{a}^{N-1}\phi^{a}Q^{a}]$ in the sense that in the
final expression only $Q^{a}$ belonging to Cartan subgroup enter. After
integration over $\nu$ one finds
\begin{equation}
\delta (Q^{a}) = \int d \mu(\phi) \exp [i \sum_{a}^{N-1}\phi^{a}Q^{a}] =
P^{0}.
\label{48}
\end{equation}
\noindent
The last step we need to do is to introduce a summing over eigenvalues
of the $Q^{a}$ of the Cartan subgroup. Just eigenfunctions of these
operators and eigenfunctions of the Hamiltonian produce a full set of the
local gauge invariant states in such a treatment.

The operator $P^{r}$ at $r \neq 0$ projects onto the invariant states
in the presence of the background probe charges. To prove it we have to
consider the $\delta$-function $\delta (Q^{a}-q^{a})$ where $q^{a}$
are the probe charges and the generators of the $r$-th irreducible
representation. Summing over all charges belonging to $r$-representation
we easily find
\begin{equation}
\sum_{q^{a} \in r} \delta (Q^{a}-q^{a}) = P^{r} =
\int d \mu (\phi) \Omega^{\star ,r}(\phi)
\exp [i \sum_{a}^{N-1}\phi^{a}Q^{a}].
\label{49}
\end{equation}

Now let us consider the Wilson lattice action in the same strong
coupling approximation (restricting ourselves to chromoelectric part of the
full action in (\ref{17}) at finite temperature).
It is known that after integration over space gauge fields and taking a limit
$a_{\beta} \rightarrow 0$ in the time direction we come to the same expression
for the partition function as it appears in the Hamiltonian formulation.
Since in the last case we do not presuppose any gauge and start from
an apparently gauge invariant formulation let us  demonstrate it once more.
To integrate out the space gauge field configurations we follow the standard
procedure and expand the Wilson action into series over the characters
of the irreducible representations of $SU(N)$-gauge group. Then, performing
the invariant integration over $dU_{n}(x)$ we come to the following expression
\begin{equation}
Z_{G} = \int \prod_{x,t}d \mu [U_{0}(x,t)] \prod_{x,n} [ \sum_{r}
C_{r}^{N_{\beta}}(\beta_{\sigma}) (Sp \prod_{t=1}^{N_{\beta}}U_{0}^{r}(x,t))
(Sp \prod_{t=1}^{N_{\beta}}U_{0}^{+,r}(x+n,t)) ]
\end{equation}
\noindent
where $\beta_{\sigma} = \frac{2N_{c}a_{\sigma}}{g^{2}a_{\beta}}$.
We can now choose Weyl's representation (\ref{38}) for the Polyakov loop
$W^r = \prod_{t=1}^{N_{\beta}}U_{0}^{r}(x,t))$ rewriting the invariant measure
in the corresponding form. After cancellation of the non-diagonal parts of
the Polyakov loops we have
\begin{equation}
Z_{G} = \int \prod_{x,t}d \mu (\alpha_{x,t}) \prod_{x,n} [ \sum_{r}
C_{r}^{N_{\beta}}(\beta_{\sigma}) (Sp W^{r}(\alpha_{x}))
(Sp W^{+,r}(\alpha_{x+n})) ]
\end{equation}
\noindent
where $W_{x} = \exp [i \sum_{t=1}^{N_{\beta}}\alpha_{x}^{a}(t) T^{a}]$,
$T^{a}$ are diagonal generators of $SU(N)$. It is enough to integrate
in the last equation only over one chosen $\alpha (t^{\prime})$.
Because of the invariant integration the result will be independent of
$\alpha (t \neq t^{\prime})$. Thus, we obtain
\begin{equation}
Z_{G} = \int \prod_{x}d \mu (\alpha_{x}) \prod_{x,n} [ \sum_{r}
C_{r}^{N_{\beta}}(\beta_{\sigma}) \Omega^{r}(\alpha_{x})
\Omega^{+,r}(\alpha_{x+n}) ].
\label{50}
\end{equation}

We consider now the quark contribution to the partition function.
To get it in the Hamiltonian formulation we have to pick up the
operator $P^r$ (\ref{40}) when $r \neq 0$ and take into account
both the states without charges and the contribution of the
antiparticles when summing over $q^a$ is performed in (\ref{49}).
In such a way one finds the quark contribution to be
\begin{equation}
\prod_{x} (1 + Re \Omega_{x}^{r}).
\label{51}
\end{equation}
\noindent
The same contribution can be obtained in the Euclidean formulation
where we do not fix any gauge. In this case we consider the interaction
of matter fields with $U_{0}(x,t)$ and explore the Kogut-Susskind massless
fermions \cite{kogut}.	Calculating the partition function
\begin{equation}
Z_{q} = \int \prod_{x} d \overline{\Psi}_{x} d \Psi_{x}
\exp (- \overline{\Psi}_{x} D_{0} \Psi_{y}),
\end{equation}
\noindent
where
$$
D_{0} = \frac{1}{2} (U_{0}(x,t) \delta_{y,x+0} - U_{0}^{+}(x,t) \delta_{y,x-0})
$$
and $U_{0}$ belongs to the fundamental representation of $SU(N)$, we get
\begin{equation}
Z_{q} = \prod_{x} (1 + Re \Omega_{x}).
\label{52}
\end{equation}
\noindent
This contribution is exactly the same as in the Hamiltonian formulation
(\ref{51}) if we choose $r \in$ fundamental representation of $SU(N)$.
To do the last step in our proof we notice that at
$a_{\beta} \rightarrow 0, N_{\beta} \rightarrow \infty$ we have
$$
C_{r}^{N_{\beta}}=( \int d \mu \Omega_{r} \exp (\beta_{\sigma}
\Omega_{\frac{1}{2}}))^{N_{\beta}} \approx \exp (- \gamma C_{2}(r))
$$
up to an irrelevant constant, $\gamma = \frac{\beta g^{2}}{2a_{\sigma}}$.
Here, $C_{2}(r)$ is the quadratic Casimir operator which is, on other hand,
the eigenvalue of the lattice Hamiltonian in the strong coupling
approximation (\ref{31}). Thus we can see the full equivalence between
the lattice Hamiltonian formulation and lattice in the Euclidean space.

Two points we would like to point out once more are: 1) the integration
over space gauge field plays in the Euclidean space a role of
the projection onto local gauge invariant states;
2) just (and only) this integration or gauge
transformation carried out at projection (\ref{45}) produces
the interactions of the eigenvalues of the Polyakov loops.

Certainly, all described above is quite known in the context of the lattice
gauge theories. Reminding this mathematical picture we wanted to make
the readers sure of two facts: 1) the lattice Hamiltonian formulation is quite
reliable approach for calculation of the effective action for the
eigenvalues of the Polyakov loop; 2) relying on apparent gauge invariance
of the lattice approximation on the whole we can claim that if
$<A_{0}> \neq 0$ appears it will be a gauge-invariant phenomenon.
This question
is a rather important one and it is, basically, the central point of the
loop calculations presented in the next chapter.

Now we are able to consider the problem of constructing of the effective
action for the eigenvalues of the Polyakov loop. At first we consider pure
gluodynamics. The quark contribution is calculated in the chapter 7.
The first calculations of the condensate on the lattice were presented in
\cite{polonyi} where the authors carried out the corresponding Monte-Carlo
computations for $SU(3)$ gluodynamics. Although the computations were done
on the small lattice the unambiguous result was obtained: $<A_{0}> \neq 0$
in the deconfined phase and the condensate falls at the deconfinement
phase transition. It signifies that $Z(N)$ and $SU(N)$ global symmetries
are broken at the same critical temperature.
Lastly, the similar result comes from
\cite{ogilv} where a simulation of the pure gauge theory fixing the Landau
gauge has been carried out. They have argued that $A_0$ develops a
nonvanishing expectation value above the deconfinement transition
temperature and have found the breaking of the colour charge
conjugation symmetry.

Unfortunately, we do not know any other attempts to compute $<A_0>$ by
means of the MC-simulations. Here, we are going to explore some known
analytical methods of statistical mechanics and spin systems. The following
material is founded on the papers \cite{bor}, \cite{bogolyubov},
\cite{survey}, \cite{aopl} where all omitted (especially technical)
details can be found.

Our starting point is the $SU(N)$ partition function (\ref{42}), (\ref{43}),
with the projection operator (\ref{44}). We define the effective
action in the standard way as \cite{wipf}, \cite{pisar}
\begin{equation}
S_{eff}(<\beta gA_0>=\phi) = - \ln Z_{\phi}(\beta ,g)
\label{53}
\end{equation}
\noindent
where
\begin{equation}
Z_{\phi} = SpK_{\phi}^0 \exp (-\beta H)
\label{54}
\end{equation}
\noindent
\begin{eqnarray}
K_{\phi}^0 = \int \prod_{x,\nu} dU_{\nu}(x) U_{\nu}(x)
\int \prod_{x}d \mu (\phi_{x}) \prod_{x,\nu}
\Omega (\phi_{x}) U_{x,\nu}^{+,l} \Omega^+ (\phi_{x+\nu}) \nonumber   \\
\delta (N\phi - \sum_{x} \phi_x),
\label{55}
\end{eqnarray}
\noindent
and $N$ is the number of the lattice sites in the spatial directions.
In these terms the operator (\ref{46}) becomes after substitution
$\phi_x \rightarrow \phi + \alpha_x$
\begin{equation}
P = \prod_{x}P_{x}^{0}, \
P_{x}^{0} = \int d \mu (\phi_{x}) \exp (-iQ^{a}_{x}
(\phi + \alpha_{x})^{a}).
\label{56}
\end{equation}
\noindent
Integration in (\ref{54}) is performed over compact measure with
the constraint
\begin{equation}
\sum_{x} \alpha_{x} = 0.
\label{57}
\end{equation}
\noindent
 From the last two equations we can deduce that the operator (\ref{56})
projects onto local gauge-invariant states with non-zero global colour charge.
In a sense it clarifies the physical nature of $A_0$-condensate: it is
an imaginary chemical potential for the colour charge out of $SU(N)$ group.
The simplest case when $<A_0> \neq 0$ may appear is the one of
$SU(2)/Z(2)$ gauge group (since elements of center are absent in such
a theory $<A_0> = 0$ in the unbroken phase). Let us consider in brief this
theory as a simple but non-trivial example which demonstrates apparently
how the condensate appears in obvious manner. The initial action
(\ref{50}) for $SU(2)/Z(2)$ theory includes the adjoint representation
for gauge field matrix $U_n(x)$. It means that we must use the sum
in (\ref{54})
only over representations which transform trivially under $Z(N)$ rotations.
Performing the general steps sketched above and summing over all
irreducible representations we obtain $Z_{\phi}$ in the form
\begin{equation}
Z_{\phi}^{SU(2)/Z(2)} = \int_{0}^{\pi} \prod_{x}d \phi_x
[\sin \phi_x]^{2(1-d)} \prod_{x,n} [\Theta_2(e^{-\gamma},\phi_x - \phi_{x+n})
- \Theta_2(e^{-\gamma},\phi_x + \phi_{x+n})]
\label{58}
\end{equation}
\noindent
where $\Theta_2$ is the Jacobi function. Making use the obvious approximation
$\phi_x \approx \phi + \delta \alpha_x$ and neglecting all fluctuations
$\delta \alpha_x$ (as the fluctuations are small they can not influence
the final result because for the gaussian free fields the constraint
$<\delta \alpha_x> =0$ is automatically fulfilled; of course, it may be not
obvious and one should argue this point by calculating the corrections;
we shall mention this point below for more realistic $SU(2)$ group)
one finds the effective action to be
\begin{equation}
-S_{eff}(\phi) = d \ln [1-\frac{\Theta_2(e^{-\gamma},2\phi)}{f(\gamma)}]
+ (1-d) \ln \sin^{2} \phi.
\label{59}
\end{equation}
\noindent
We omitted an irrelevant constant and $f(\gamma ) = \sum_{l=0,1,..}^{\infty}
\exp (- \gamma l(l+1))$. Analyzing this expression we conclude that there
exists such $\gamma_c$ that at $\gamma > \gamma_c$, $\phi neq 0$
in the point of minimum of the effective action (\ref{59}).
This conclusion is quite understandable without the approximation we have
used here. The Jacobi function in (\ref{58}) which carries the sign
"$+$" does not include the constant part of $\phi_x$. This constant part
enters in the second Jacobi function with sign "$-$". As $\phi \neq 0$
lowers the Jacobi function it signifies that when $\phi \neq 0$, $Z_{\phi}$
increases and, thus, leads to deeper minimum of the effective action
than $\phi =0$. Unfortunately, such (almost "strict") proof  can be done
only in this simple example. The more realistic case of $SU(2)$ gauge group
has been investigated in \cite{bor}. As this case will be presented below
and in the more general context we restrict ourselves to some phrases here.
We used the same approximation as described above  and calculated the
corrections coming from Gaussian integration over $\delta \alpha_x$.
The obtained $S_{eff}$ was investigated numerically. Our conclusions are
similar to the previous case.

Lastly, we are studying QCD ($SU(3)$ gauge group) in details by two
independent methods of calculations in order to be convinced in the results.
First, we suppose
that all fluctuations are suppressed by coupling constant. Recalling that
$SU(3)$ representations are labelled by two independent indices
$l=(l_1, l_2)$ and eigenvalues of $E^2$ are the quadratic Casimir operator
eigenvalues $C_2(l)=\frac{1}{3}(l_1^2+l_2^2-l_1l^2) -1$ we can perform
a summation in the partition function (\ref{54}) over all irreducible
representations. In this way we can calculate the single-site free energy
in the form convenient for the following analysis
\begin{equation}
-S_{effG}^{SU(3)}= \ln [F(\phi)]= \ln [\frac{\Xi^d(\phi)}{\mu^{d-1}(\phi)}]
\label{60}
\end{equation}
\noindent
where for readability we have introduced the notations
\begin{equation}
\mu (\phi)= \sin^2(\frac{\phi_1-\phi_2}{2}) \sin^2(\frac{2\phi_1+\phi_2}{2})
\sin^2(\frac{2\phi_2+\phi_1}{2})
\label{61}
\end{equation}
\noindent
and
\begin{eqnarray}
\Xi=\exp (\gamma) \sum_{\alpha =1}^3 [-\frac{1}{3}
\Theta_3(\gamma /3,0)\Theta_3(\gamma, 2\Phi_{\alpha}) -\frac{2}{3}
\Theta_3(\gamma /3,\Phi_{\alpha})\Theta_3(\gamma,\Phi_{\alpha}) + \nonumber \\
\frac{1}{3}\Theta_3(\gamma /3,\Phi_{\alpha}) \sum_{\beta > \sigma}
\Theta_3(\gamma /3,\Phi_{\beta}-\Phi_{\sigma})] +[\Theta_3 \rightarrow
\Theta_2]
\label{62}
\end{eqnarray}
\noindent
up to an irrelevant constant. Here $\Theta_i(\gamma, \phi)$ are the Jacobi
functions and $\Phi_{\alpha} = \phi_{\beta}-\phi_{\sigma}$ at
$\sum_{\alpha}\Phi_{\alpha}=0$ ($\phi_{\alpha}$ has been defined after
eq.(\ref{27})).
Thus $<\phi>$ can be found from the local minima of $S_{effG}$.
The $S_{effG}$ behaviour  has been analyzed numerically at $d=3$ and in the
interval of $\gamma = [1,2]$ with high precision.
At larger $\gamma > \gamma_c$
(small temperature) the minima of $\ln [F(\phi)]$ are located inside of
the every triangle plotted on Fig.1 at the following values:
$<\phi_1>=2\pi k, <\phi_2>=2\pi (k+\frac{1}{3}), <\phi_3>=2\pi (k-\frac{1}{3})$
and so on (six combinations are possible).
Obviously, that this distribution is invariant under the
$Z(3)$ transformations and we conclude that the system is in the
confined phase
where $<TrW>=0$ and $<\phi>=0$. With $\gamma$ decreasing (temperature
increasing) at $\gamma=1.75$ three secondary minima of the function
$\ln [F(\phi)]$  develop and get deeper. At the same time the initial
minimum of $\ln [F(\phi)]$ gets also deeper but it develops slower
and at $\gamma =1.62$ it disappears at all. But still earlier at
$\gamma_c^G=1.73$ the secondary minima become degenerate with the initial one
and the system could undergo the phase transition presumably of the first
order. In these newly developed minima $<\phi> \neq 0$
which signals the global
gauge symmetry breakdown and  forming  temperature dependent condensate.
It is clear that all these phenomena emerge at the deconfinement
phase transition since above its critical temperature $<TrW> \neq 0$ in any
minimum.

The second approach to evaluate the condensate for the $SU(3)$ gluodynamics
has been presented in \cite{bogolyubov}. To understand the picture in more
details authors of \cite{bogolyubov} have considered the first non-trivial
term in sum over characters and reparametrized the conditions
$\delta (N\phi^a - \sum_{x}\phi^a_x )$ in terms of the Polyakov loop.
For calculation of the effective action Bogolyubov's method of
quasiaveraging has been utilized. Numerical studying of the effective
action up to $\lambda^{12}$ (where $\lambda$ is an effective
coupling constant which is small in the strong coupling region) shows that
the condensate appears
in the deconfinement phase and moreover the picture described above is
approximately reproduced in the framework of Bogolyubov's method
and of redefinition
of those quantities over which we have to minimize the effective action.
In a sense, this method is close to those developed in \cite{belyev3},
\cite{bhat}. The difference is that we used the Polyakov loop to express
the effective action (though in a special parametrization) from the very
beginning. Our final result is directly opposite to the one obtained
in \cite{belyev3}, \cite{bhat}. It might imply that Belyaev's method is not
self-consistent (a comprehensive analysis of this method we give in the next
chapter).

Generally speaking, the situation could turn out to be more complicated than
described above. At high temperature the Polyakov loop
develops a non-vanishing expectation value.
In quantum theory, the Polyakov loop, being the function of the random
variable $A_0$, is an other random variable. It could mean that
to obtain the real distribution of contributions to the free energy coming
from the Polyakov loop and from $A_0$-condensate we need to construct
a general
effective action for both these quantities and to find minima of such an
effective action. In other case it cannot be excluded that either
some part of the condensate may be "transfered" into the Polyakov loop
or vice versa.
To verify this idea we carried out the corresponding calculation for
$SU(2)$ gauge group. Our preliminary results are the following \cite{aopl}.
Implanting $\delta$-function $\delta (N\sigma -\sum_{x}\cos \phi_x)$ in
(\ref{55}) we considered fundamental $SU(2)$ representation
in the sum over characters
and used the standard mean-field approximation to calculate
the effective action $S_{eff}(\sigma, \phi)$. Then, minimizing
$S_{eff}(\sigma, \phi)$ we found that both $\sigma$ and $\phi$ differ
from zero in the point of minimum of the $S_{eff}(\sigma, \phi)$ in the
deconfinement phase. Hence, all the conclusions derived above have been
confirmed in this approach. It seems to us that in compact theories like
lattice gauge theory we should use just this scheme of calculations, because
invariant integration can, in fact, produce the same $<\alpha> \neq 0$
as $<\phi> \neq 0$ (see (\ref{54})) but entering with opposite sign and, so,
cancel the condensate. It may be not the case in the loop expansion method,
for at $g^2 \rightarrow 0$ the constraint on $\alpha$ in (\ref{54}) is
automatically fulfilled for the gaussian fluctuations.

We have, thus, found out that the chromoelectric sector of
the lattice gauge theory
generates $A_0$-condensate in the deconfinement phase. In all considered
approaches the obtained picture is essentially the same, perhaps
up to irrelevant details.
Since the field $A_0$ and the relevant Polyakov loop transform
non-trivially under the centre subgroup transformations the condensate
carries out the charges of the centre (as matter fields). This leads
to quark colour charge screening and vanishing the long-range forces
which explain, in fact, the deconfinement mechanism.

Concluding this chapter we would like to make some remarks on
Ref.\cite{pisar}. The authors of the paper have considered
a partition function for the eigenvalues of the Polyakov loop in the
continuum theory. Their consideration is very close to ours described here
as they have used the temporal gauge $A_0=0$ with projection onto the gauge
invariant states. The conclusion of the paper is that there is no
real condensation at high temperature. The basic assumption
conjectured by the authors is the cancellation of the Vandermonde
determinant. We, however, reckon that the real situation with the cancellation
is a little more complicated than described in \cite{pisar}.
The possibility of the cancellation of the Vandermonde
determinant (group integration measure) does exist.
It can appear just after integration over spatial gauge field, which
one may see from (\ref{50}) if we consider a contribution
only of spatially longitudinal mode. The partition function has a form of
the character expansion. The character representation can be chosen
in such a form that denominator of  the sum over characters will
include group integration measure. And, of course, in this sense
the cancellation takes place. Let us now return to
the Hamiltonian version where we must use a projection operator
onto gauge invariant states in gauge $A_0=0$. It is known for a long
time (and we stressed this point several times in this chapter)
that integration over spatial gauge fields is equivalent in a sense
to projection onto gauge-invariant states in the gauge $A_0=0$.
We have shown above that it leads to the same character expansion.
In the Hamiltonian formulation the character expansion results
from  projection
operator and summing over diagonal group operators which commute with the
Hamiltonian (the last point was missed in the discussed paper).
It means the following: Choosing the same representation
for $SU(N)$ characters we can formally cancel the determinant.
But on the other hand, it means that we cancel some part of the projection
operator. The sum over irreducible $SU(N)$ representations which
survives after the cancellation can never be presented as the old
projection operator (it can be easily demonstrated mathematically).
The inconsistency of Ref.\cite{pisar} is
the usage of the same projection operator after the cancellation.
Strictly speaking, this is incorrect. Namely this gives a possibility
to rewrite the partition function in such a form that the constant part of
$A_0$ will be only at imaginary unit. Thus, there are two
possibilities: either one cancels the determinant but after that
we must build a new projection operator (if it can exist)
or we can work with the usual projection operator but the determinant
is not canceled in this case.
In chapter 6 we shall give a confirmation what we sketched above of.
The same determinant
appears not only in the Hamiltonian version but also in the static
gauge for $A_0$  as the Fadeev-Popov determinant. There is no doubt
that cancellation takes place and besides not only at one-loop
level. But nevertheless the condensate falls at high temperature.
It may signify only one fact: after the cancellation we have no
old simple representation for the projection operator.

Let us consider in the static gauge the part of the Yang-Mills action
quadratic in spatial fields.
Basically, the term at linear power in
$A_0$ is nothing but Gauss' law. Integration over
$A_0$ gives as a result  the projection operator onto the states
where Gauss' law is fulfilled. Let us now make an integration over
spatial fields. Since we have limited ourselves to quadratic part,
this integration can be performed exactly.
Resulting determinant appears to be the Vandermonde
determinant but in $(-1/2)$ power as usual. In such a way we come
to the obvious cancellation. We verified all these results
in finite-difference
lattice formulation in the same static gauge \cite{bbpznew}.
Then, in order to come to the partition function \cite{pisar}
we must look for such an expression for the projection operator
where $A_0$ enters again at Gauss' law in the action. Only then
we will be able to separate the constant part  of $A_0$ in the exponent.
It is impossible, which of course can be easily seen from the formulae
of this chapter.

Probably, a more appropriate partition function in the continuum
theory just for $A_0$ gauge field has been conjectured in \cite{poljon}
where the authors have introduced an invariant integration over $A_0$
and postulated an effective action for this variable. This theory
is able to describe confinement in QCD. We think that theory
can exhibit spontaneous symmetry breaking by means of $A_0$ field
condensation at high temperature \cite{bbpznew}.

\section{$<A_0>$ in loop expansion}.

First, a possibility of $A_0$-condensation had been discussed
in the standard loop expansion approach to the calculation
of the effective action \cite{gross}, \cite{anish}, \cite{belyev}.
Just in this method the most essential questions such as gauge invariance
of the condensate, the higher loop contributions, the thermodynamics
of the $Z(N)$-phases have been examined. In this chapter we are going
to go in a systematic way through the results of the calculations and
to discuss them. Our actual calculations  will be done for QCD in the limit
of high temperatures when the coupling constant is small and it is possible
to expand in this parameter. All calculations are carried out in the
background relativistic gauge $R_{\xi}^{ext}$ which incorporates the results
of all relativistic gauges.

We begin with calculation of the two-loop effective action
$W^{(2)}(A_{0}, \xi)$ of the gluonic external field due to the pure gluon
contribution  \cite{belyev}, \cite{anish}, \cite{enq}, \cite{scl8},
\cite{scl2} (the quark contribution will be analyzed in chapter 7).
The QCD Lagrangian in the relativistic background gauge reads:
\begin{eqnarray}
      {\cal L}&=&\frac{1}{4}(G^a_{\mu\nu})^2+\frac{1}{2\xi}
      (D^B_\mu Q^a_\mu)^2
      + \bar\chi D^B_\mu D_\mu \chi			      \nonumber \\
      &&+ \bar\Psi^a(\gamma_\mu\partial_\mu +im)\Psi^a +
      ig \bar\Psi^a\gamma_\mu({ A}^c_\mu + Q^c_\mu)
      (t^c)^a_b\Psi^b , 						\\
      G^a_{\mu\nu}&=&(D^B_\mu)^{ab}Q^b_\nu -
      (D^B_\nu)^{ab}Q^b_\mu -gf^{abc}Q^b_\mu Q^c_\nu,
							      \nonumber \\
      (D^B_\mu)^{ab}&=& \delta^{ab}\partial_\mu + gf^{abc}{ A}^c_\mu, \
      (D_\mu)^{ab}= \delta^{ab}\partial_\mu + gf^{abc}
      (Q^c_{\mu}+A^c{\mu}),				      \nonumber \\
      { A}^c_\mu & = &\delta_{\mu 0}(\delta^{c3}{ A}^3_0
      + \delta^{c8}{ A}^8_0),				      \nonumber
\label{lct}
\end{eqnarray}
\noindent
where $(t^c)^a_b$ are the $SU(3)$ generators, $Q^a_\mu$ is the quantized
field, $f^{abc}$ are the structure constants and $\bar{\chi}, \chi$
are the ghost fields.
In what follows it will be convenient to introduce the "charged basis"
of the gluonic fields:
\begin{eqnarray}
      &&\pi^\pm_\mu = \frac{1}{\sqrt{2}}(A^1_\mu\pm i A^2_\mu)~~~~,~~~~
      \pi^0_\mu = A^3_\mu~~~~,~~~~
      K^\pm_\mu = \frac{1}{\sqrt{2}}(A^4_\mu\pm i A^5_\mu)~~~~,~~~~
       \nonumber   \\
      &&{\bar K}^\pm_\mu =
      \frac{1}{\sqrt{2}}(A^6_\mu\pm i A^7_\mu)~~~~,~~~~
      \eta_\mu = A^8_\mu .
\label{64}
\end{eqnarray}
In this basis the background fields $A^{3}_{0}$ and $A^{8}_{0}$
enter into the momentum space Lagrangian as constant shifts of the
zero momentum components. More details about the "charged basis"
(\ref{64}) and the Feynman rules are given in \cite{belyev}.

To determine the vacuum  value	of  $A_0$  one should calculate  the
effective action $W(A_0,\xi)$ and find its minimum point $(A_0)_{min}$
via the minimization procedure. If  this  value   occurs
to  be different  from $\frac{2 \pi n}{\beta g}$, it means that  spontaneous
breaking  of  the gauge symmetry happens.

The effective action $W(A_0,\xi)$ is given as a functional integral
over periodic gauge and ghost fields and antiperiodic fermion fields
(\ref{22})
\begin{equation}
\exp [-W(A_0)VT]= N \int{\cal D}Q{\cal D}\bar\Psi{\cal D}\Psi
{\cal D}\bar\chi{\cal D}\chi
e^{-\int_{0}^{\beta}d^3x({\cal L}+J^a_\mu Q^a_\mu)},
\label{65}
\end{equation}
\noindent
where $\cal{L}$ is the Lagrangian (\ref{lct}), $N$ is a $T$-independent
normalization factor, $V$ is space volume and
$J^a_\mu$ is an external source. The effective action due to gluons
up to two-loop order in the background gauges has been calculated
in Ref. \cite{anish}, \cite{belyev}, \cite{enq}, \cite{scl8},
\cite{scl2}, \cite{chub}, \cite{sclnew}, \cite{kalnew}, \cite{saw}.
Since the result has crucial significance for what follows we have
calculated it once more. Our result for $W(A_0, \xi)$ is the following
$$
W(A_0, \xi)\beta^{4}=\frac{4\pi^2}{3}(-\frac{1}{30}
    +\sum^{3}_{i=1}B_4(a_i) )
+\frac{g^2}{2} [\sum^{3}_{i=1}[ B^2_2(a_i)+2B_2(0)B_2(a_i) ]
+B_2(a_1)B_2(a_2)	\nonumber  \\
$$
\noindent
$$
+ B_2(a_2)B_2(a_3) + B_2(a_3)B_2(a_1) ]
+\frac{1-\xi}{3}g^2
[ B_3(a_1) [2B_1(a_1)+B_1(a_2)-B_1(a_3) ]	      \nonumber  \\
$$
\begin{equation}
 +B_3(a_2)[2B_1(a_2) + B_1(a_1)+B_1(a_3)]
+B_3(a_3) [2B_1(a_3)+B_1(a_2)-B_1(a_1)]
\label{hrju}
\end{equation}
\noindent
where the notations have been introduced
\begin{eqnarray}
x=\frac{g\beta A_0^3}{\pi}, \ y=\frac{g\beta A_0^8}{\pi}; \nonumber   \\
\ a_1=\frac{x}{2}, \ a_2=\frac{1}{4}(x+\sqrt{3}y), \
a_3=\frac{1}{4}(-x+\sqrt{3}y),
\label{67}
\end{eqnarray}
\noindent
This expression  differs  from	that  of ref. \cite{enq}  in the $\xi$-
dependent part by the additional factor 3/2 . Besides,
all signs in the squared brackets are "plus" in \cite{enq}.
Our result (\ref{hrju}) coincides (up to the sign definition)
with the result of ref.\cite{kalnew}. As is seen
from (\ref{hrju}), the gluon  contribution
depends on the gauge fixing parameter  $\xi$.
In QCD at high temperature the coupling constant is small.
So,  one  can  calculate the minimum point of $W(A_0,\xi)$  and
the value of the functional in the minimum by an  expansion  in $g^2$.
Up to the second order we obtain:
\begin{equation}
\beta^4 W_{gl}=-\frac{8}{45}\pi^2 + \frac{1}{6}g^2 -
\frac{g^4}{32 \pi^2}(3-\xi)^2	      \nonumber    \\
x_{min}=\frac{g^2}{4 \pi^2}(3-\xi), \  y_{min} = 0
\label{68}
\end{equation}
\noindent
Here, the values of $x_{min}$ and $y_{min}$ have been found for the
intervals $0 \leq a_{1,2} \leq 1, -1 \leq a_3 \leq 0$.
Five other minima in the $(x,y)$-plane can be found by	means  of
consequent rotations of  the coordinate system by the angle
$\frac{\pi}{3}$. From (\ref{68}) it follows that the presence
of the condensate lowers the action. So, the spontaneous generation
of the condensate takes place. It is very essential that the depth
of all minima is the same. Hence it follows that although the condensate
breaks both gauge and $Z(3)$ global symmetries, the generated phases
preserve the rotation symmetry as in the case of $Z(N)$-phases. The
description of the general structure will be done below after
evaluating the quark contribution. These results are in the full
accordance with those from the previous chapter. Nevertheless it is to be
emphasized that in the loop expansion treatment there is no phase
boundary whereas in the strong coupling lattice approach there	is a
phase transition to the state with $<A_0> \neq 0$.

As is also seen from
(\ref{68}), the vacuum value $x_{min}$ as well as the minimum
value of the functional $W(x_{min})$ turn out to be dependent on
the gauge fixing parameter $\xi$. This is the $\xi$-dependence problem
discussed by many authors \cite{belyev3}, \cite{enq}, \cite{scl8},
\cite{scl2}, \cite{sclnew}, \cite{kalnew}, \cite{bhat}.
In general, in gauge theories there are two ways of dealing with
the problem of gauge invariance: 1)to use the explicitly gauge invariant
formulation from the very beginning; 2) to apply the Green function
analysis and extract the gauge invariant results for the observables
via the Nielsen (the Ward type) identities. Besides, some hybrid method
can be developed as well. As far as the problem of the $A_0$-condensate
is concerned,
methods of all kind have been applied and different conclusions have been
extracted. In the previous chapter the explicitly gauge invariant approach
to $A_0$ was presented. In that scheme $<A_0> \neq 0$ is obviously
a gauge invariant phenomenon. Here, we are going to explore the Nilson
identity to demonstrate gauge invariance of the condensate.

One of the powerful method of dealing with the $\xi$-dependence problem is
to apply the Nielsen identities method which first was used in the
investigations of spontaneous gauge symmetry breaking
by radiative corrections \cite{niels}, \cite{kugo}, \cite{fraser}. As is well
known, in models of the Coleman-Weinberg type both the effective potential
$V(\Phi, \xi)$ of the scalar field $\Phi$ and its minimum point
$\Phi_{min}$ are $\xi$-dependent. A very elegant resolution of this
problem had been found by Nielsen \cite{niels} who proved the gauge invariance
of this dynamical phenomenon. Namely, he formulated the Ward type identity
describing an analytical dependence of $V(\Phi , \xi)$ on $\xi$. So any
variation of the potential can be compensated by the corresponding variation
in $\delta \Phi$ (along some characteristic line in the $(\Phi, \xi)$-plane).
More details about these results can be found in \cite{niels}, \cite{kugo},
\cite{fraser}. In the papers \cite{scl8}, \cite{scl2}, \cite{sclnew} this
method has been applied to the problem of $\xi$-dependence of the gluon
condensate. First of all this gives a possibility to check the correctness
of the calculations. Secondly, due to generality  of the approach
the relation to other methods of calculations can be established. Nielsen's
identities of a general form have been recently derived by Kobes,
Kunstatter and Rebhan \cite{rebhan}. For the effective action
they describe a variation of $W(\overline{\Phi})$ due to a variation
of the gauge fixing term $F^{\alpha}(\Phi)$ and in the condensed De Witt
notations are given by the expression:
\begin{equation}
\delta W(\bar{\Phi}) = W_{,j} \delta \Xi^{j}(\bar{\Phi}).
\label{69}
\end{equation}
\noindent
Here $\Phi^{i}$ is a gauge field, superscript "i" includes all discrete
and continuous variables, $\overline{\Phi}^{i}$ denotes a vacuum value
of the field, comma after $W$ means the variation derivative with respect
to corresponding fields and the contraction means integration over
continuous and summation over discrete variables. The variation $\delta
\Xi^{i}$ describes changing of the mean field value due to the special
gauge transformations
\begin{equation}
\delta \Phi^i = D^i_{\alpha}(\Phi) \delta \Omega^{\alpha}
\label{70}
\end{equation}
\noindent
with the parameter
\begin{equation}
\delta \Omega^{\alpha} =- \Delta^{\alpha}_{\beta}(\Phi) \delta^{\prime}
F^{\beta}(\Phi).
\label{71}
\end{equation}
\noindent
This parameter is chosen to cancel the change  $\delta S$ of the classical
action
\begin{equation}
S_{g.f.}(\Phi) = S(\Phi) + \frac{1}{2}\eta_{\alpha , \beta} F^{\alpha}(\Phi)
F^{\beta}(\Phi)
\label{72}
\end{equation}
\noindent
due to the variation of the gauge fixing term, $\delta F^{\alpha}(\Phi)$,
and the "metric" $\eta_{\alpha, \beta} \rightarrow \eta_{\alpha, \beta} +
\delta \eta_{\alpha, \beta}$:
\begin{equation}
\delta^{\prime}F^{\beta} = \delta F^{\beta} + \frac{1}{2}\eta^{\rho, \beta}
\eta_{\alpha, \rho} F^{\alpha}
\label{73}
\end{equation}
\noindent
In Eqs.(\ref{70}-\ref{73}) $D^{i}_{\alpha}(\Phi)$ are gauge
transformation generators, $\Delta_{\beta}^{\alpha}(\Phi)$ is the
ghost propagator in the external field $\Phi$, $S(\Phi)$  is the
classical action of the gauge fields. In quantum theory $\delta \Xi^{i}$
has to be calculated from the equation \cite{rebhan}
\begin{equation}
\delta \chi^i = -<D^{i}_{\alpha}(\Phi) \Delta_{\beta}^{\alpha}(\Phi)
\delta^{\prime} F^{\beta}(\Phi)>
\label{74}
\end{equation}
\noindent
where
\begin{equation}
<O(\Phi)> = e^{-W(\Phi)} \int D \Phi  O(\Phi)
det[F_{,i}^{\alpha}(\Phi)D^i_{\beta}]
\exp (-S_{g.f.} -W_{,j}(\bar{\Phi})(\Phi - \bar{\Phi})^j )
\label{75}
\end{equation}
\noindent
and we substituted the current $J_{j} = -W_{,j}$.  In the background
field gauge the gauge fixing functions depend on an additional external
field $\tilde{\Phi}^i$. So the expression (\ref{74}) should be replaced
by
\begin{equation}
\delta^{\prime} \chi^{i} = (C_j^i)^{-1} \delta \chi^j
\label{76}
\end{equation}
\noindent
where the function
\begin{eqnarray}
C_j^i(\bar{\Phi}) = \delta_j^i - <D^i_{\alpha}(\Phi) \Delta_{\beta}^{\alpha}
(\Phi)(\frac{\delta F^{\beta}(\tilde{\Phi},\Phi)}
{\delta \tilde{\Phi}^j}+  \nonumber   \\
\frac{1}{2}\eta^{\beta \gamma}\frac{\delta \eta_{\gamma \rho}}
{\delta \tilde{\Phi}^j}F^{\rho}(\Phi ,\tilde{\Phi})>_{\tilde{\Phi}=\bar{\Phi}}
\label{77}
\end{eqnarray}
\noindent
describes the dependence of the function $F^{\alpha}$ on $\tilde{\Phi}$.  The
additional external field must be set equal to the vacuum value
$\overline{\Phi}^i$ at the last step of the calculations. Besides, the latter
has to be calculated from the effective action $W(\tilde{\Phi}, \bar
{\Phi})$ with $\tilde{\Phi}$  and $\bar{\Phi}$ taken to be different.
$\delta^j_i$ is the Kroneker delta. In QCD one must use the total gauge field
$A_{\mu}^{a} = Q_{\mu}^a + \delta_{\mu 0}(\delta^{a3}A_0^3 + \delta^{a8}
A_0^8)$ as the field $\Phi^i$. One must substitute
$D^i_{\alpha} = D_{\mu}^{ab}(A+Q)$ as generators and
$\eta^{\alpha \beta} = \frac{1}{\xi} \delta^{ab}$ as the metric tensor.
Variations of $\xi$ can be realized by the variations of the metric
\begin{equation}
\delta \eta^{\alpha \beta} = \delta^{\prime}F^{\alpha} = -\frac{1}{2}
(D_{\mu}^B(A)Q_{\mu})^a \frac{\delta \xi}{\xi}.
\label{78}
\end{equation}
\noindent
In the papers \cite{scl8}, \cite{scl2} the relations (\ref{76}),
(\ref{77}) have been adopted to the case of the background field gauge.
To do it the external field $\tilde{\Phi} = \overline{A}$ appearing in
the gauge fixing function $F^{\alpha}(\tilde{\Phi},\Phi)=(D_{\mu}^B
(\overline{A})Q_{\mu})^a$ should be initially assumed to be different
from the actual background field $B^a \equiv A^a = const$. The former
must be identified with the latter one at the end of the calculations.
So the field $\overline{A}$ in the covariant derivative $D^B_{\mu}
(\overline{A})$ can be written as $\overline{A}^a_{\mu} = (B+q)^a_{\mu}$
where $q^a_{\mu}$ are the deviations from the solution $B^a_{\mu}$. Taking
$q^a_{\mu}$ to be small one may calculate propagators, vertices, etc
as series in $q^a_{\mu}$ and keep only the linear terms. Then the
functions $\delta \chi^{\prime{i}}$ (\ref{76}) and $C^i_j$ (\ref{77})
are calculated by differentiating with respect to $q^a_{\mu}$. On this
way the identity (\ref{69}) can be written as:
\begin{equation}
\delta W(B)=W_{,q}(B,q) \mid_{q=0}[C^a_{\beta}(B,q)]^{-1}_{q=0}
\delta \chi^{\beta}(B,q) \mid_{q=0}
\label{79}
\end{equation}
\noindent
where
\begin{equation}
\delta \chi^{\beta}(B)=-\frac{1}{2} \frac{\delta \xi}{\xi}
<D^B_d(B+Q) \Delta^d_e(B+q) (D^B_{\mu}(B+q)Q^e_{\mu})_{q=0}>,
\label{80}
\end{equation}
\noindent
$$
C^a_{\beta}=\delta^a_{\beta} + g f^{abc} \delta_{\mu 0}\delta^b_{\beta}
<D^a_e(B+Q) \Delta^e_d(B+q) Q^c_{\mu}>_{q=0}.	\nonumber   \\
$$
The average values in (\ref{74}), (\ref{80}) should be calculated
in perturbation theory considering quantum fluctuations $Q^a_{\mu}$
and deviations $q^a_{\mu}$ to be small. The internal index "$a$"
in eq. (\ref{79}) takes now the values $a=3,8$ in accordance
with the structure of the background field.

As is seen from eq. (\ref{hrju}), the dependence of the effective
action on $\xi$ appears in the order $g^2$. So, in the lowest order both
the  left-hand	side (LHS)  and the right-hand side (RHS) of eq.(\ref{79})
should be of the order $g^2$, as well. The covariant derivative $D^a_{\mu}
(\overline{A}+Q)$ contains the quantum field as a product $gQ^a_{\mu}$.
So, in the one-loop approximation the variation $\delta \chi^{(1)}$
is of the order $g$. The one-loop effective action $W^{(1)}(B)$ has the
zero order and its derivative with respect to $B$ has the first order in
coupling constant. Hence it follows that in the  lowest order one must use
in the identity (\ref{79}) the one-loop  functions $\delta \chi^{(1)}$
and $C_{\beta}^{(1) \alpha}$ to be equal to unity.

To calculate $\delta{\chi^a}$ (\ref{80}) it  is necessary
to take into account the explicit form of generators and the fact that only
the third and eighth components of external
field are to be non-zero. Thus, in the one-loop approximation the
expression (\ref{80}) reads
\begin{equation}
\delta \chi^{(1)a}=-\frac{1}{2}g \frac{\delta \xi}{\xi} f^{abc}
<\Delta^{bd}(x-y) \tilde{D}^{dl}_{\nu}(B)G^{lc}_{\nu 0}(x-y)>.
\label{81}
\end{equation}
\noindent
The necessary structure constants in the basis (\ref{64}) are
\begin{equation}
f^{3\pi^+ \pi^-}=i, \ f^{3k^+k^-}=-f^{3\bar{k}^+\bar{k}^-}=\frac{i}{2}, \
f^{8k^+k^-}=f^{8\bar{k}^+\bar{k}^-}=\frac{i\sqrt{3}}{2}
\label{82}
\end{equation}
\noindent
and the background covariant derivatives may be written as follows
\begin{equation}
(D^B_{\mu}Q^a_{\nu})=\tilde{D}^{ab}_{\mu} \Pi^b_{\nu}
\label{83}
\end{equation}
\noindent
where $\Pi^b_{\nu}$ is the column
\begin{equation}
\Pi^b_{\nu}=(\pi^+,\pi^-,\pi^0,k^+,k^-,\bar{k}^+,\bar{k}^-,\eta )_{\nu}^{T}
\label{84}
\end{equation}
\noindent
and values of the variables $a,b$ now are: $a,b =\pi^+,\pi^-,..., \eta$.
Diagonal elements of $\tilde{D}^{ab}_{\mu}$  are the following:
\begin{equation}
diag\tilde{D}^{ab}_{\mu}=(\tilde{D}^{a_1}_{\mu},\tilde{D}^{-a_1}_{\mu},
\partial_{\mu},\tilde{D}^{a_2}_{\mu},\tilde{D}^{-a_2}_{\mu},
\tilde{D}^{a_3}_{\mu},\tilde{D}^{-a_3}_{\mu}, \partial_{\mu})
\label{85}
\end{equation}
\noindent
where $a_{i}$ stand for  the  background  fields (\ref{67})
describing the contributions of the isospins  $I$, $V$
and $U$ spin subgroups of the SU(3) group, respectively. Using
the explicit forms of the gluon and ghost field propagators
\begin{eqnarray}
G^{ab}_{\mu \nu}(x-y)=\frac{1}{\beta}\sum_{n=-\infty}^{\infty}
\int \frac{d^3p}{(2\pi)^3} e^{-ip(x-y)}  \nonumber  \\
(\delta_{\mu \nu}\frac{\delta_{ab}}{(p^a)^2} + (\xi -1)
\frac{p^a_{\mu}p^b_{\nu}}{(p^a)^4})		     \\
\delta^{ab}(x-y)=\frac{1}{\beta}\sum_{n=-\infty}^{\infty}
\int \frac{d^3p}{(2\pi)^3} e^{-ip(x-y)} \frac{\delta^{ab}}{(p^a)^2}
\label{86}
\end{eqnarray}
\noindent
where
\begin{equation}
p^e_{\mu}=[p^0+a^e,\vec{p}],  \ p^0=\frac{2 \pi n}{\beta}, n=0,\pm 1,...
\label{87}
\end{equation}
\noindent
and
\begin{equation}
a^{\pi^{\pm}}=\pm gB^3_0, \ a^{k^{\pm}}=\pm \frac{g}{2}(B^3_0+\sqrt{3}B^8_0),
\ a^{\bar{k}^{\pm}}=\pm \frac{g}{2}(-B^3_0+\sqrt{3}B^8_0), \
a^{\pi^0}=a^{\eta}=0
\label{88}
\end{equation}
\noindent
and substituting (\ref{83})-(\ref{86}) in eq.(\ref{81}) one obtains:
\begin{eqnarray}
\delta \chi^3=\frac{g}{4\pi \beta}[B_1(a_1)+\frac{1}{2}B_1(a_2)-
\frac{1}{2}B_1(a_3)] \delta \xi,   \nonumber  \\
\delta \chi^8=\frac{g}{4\pi \beta} \frac{\sqrt{3}}{2}
[B_1(a_2)+B_1(a_3)] \delta \xi.
\label{89}
\end{eqnarray}
\noindent
The derivatives of the one-loop parts of $W(A_0,\xi)$
in eq.(\ref{hrju}) with respect to $B^3=\overline{A}^3_0$ and
$B^8=\overline{A}^8_0$ equal the expressions:
\begin{eqnarray}
\frac{\partial W^1_{gl}}{\partial B^3_0}=\frac{4g\pi}{3\beta^3}
[2B_3(a_1)+B_3(a_2)-B_3(a_3)],	      \nonumber       \\
\frac{\partial W^1_{gl}}{\partial B^8_0}=\frac{4g\pi}{3\beta^3}\sqrt{3}
[B_3(a_2)+B_3(a_3)].
\label{90}
\end{eqnarray}
\noindent
By summing up the corresponding products of expressions
(\ref{89}),(\ref{90})
one obtains the gluon contribution to the RHS of eq.(\ref{79}).
The obtained expressions coincide up to the sign with
the derivative of (\ref{hrju}) with respect to $\xi$. Thus, the
Nielsen identity
\begin{equation}
\frac{d W_{gl}}{d\xi}=\frac{\partial W^2_{gl}}{\xi} +
\frac{\partial W^1_{gl}}{\partial B^3_0}C^{(1)}_3 +
\frac{\partial W^1_{gl}}{\partial B^8_0}C^{(1)}_8 =0
\label{91}
\end{equation}
\noindent
is satisfied up to the two-loop order. In eq.(\ref{91}) we have denoted
$C^{(1)}_{3,8}=\frac{\delta \chi^1_{3,8}}{\delta \xi}$.
The identity (\ref{91}) is just the characteristic equation. So,
from the equation it follows that along characteristic lines in the
$(A_0,\xi)$-plane the effective action is not changing. In particular,
this is the case for its minimum value (\ref{68}). In accordance
with general theory of Nielsen's identity approach this means that the gluon
condensation at finite temperature is a gauge invariant phenomenon. The fact
that identity (\ref{91}) holds means as well that loop expansion of
$W(A_0,\xi)$ is the self-contained procedure and no other $\xi$-dependent
diagrams should be included in the order $g^2$. Only $\xi$-independent
terms may be added, in principle, to eq.(\ref{hrju}). From Refs.
\cite{niels}, \cite{kugo} it also follows that along an orbit which passes
through the point $(\xi ,x_{min} \neq 0 )$ not only the minimum value
of the effective action $W(\xi ,(A_0)_{min})$ but all other
observables (particle masses, $S$-matrix elements, etc) are to be
constant as well. This property selects the unique orbit among other ones.

As it was mentioned before, $\xi$ dependence of the effective action
(\ref{hrju}) has called in question a possibility of the gluon
field condensation because physical phenomena  must be gauge
independent. So, in a number of papers some gauge invariant methods of
calculations have been proposed. Historically the first of them has been
used by Belyaev (\cite{belyev3}) who introduced a special reparametrization
of the background field $A_0$ in terms of the Polyakov loop (\ref{18}).
Applying this procedure to $SU(2)$ gluodynamics he came to the conclusion
that there is no real condensation at the two-loop level. The same result
has been derived by this method in $SU(N)$ gluodynamics in Ref.\cite{bhat}.
Thus, it turns out that two methods of calculations give opposite
conclusions about $A_0$-condensate (at two-loop level). Let us try to find
the origin of the discrepancy.

First of all let us briefly describe Belyaev's method and consider $SU(2)$
case for simplicity ($I$-spin subgroup of the $SU(3)$ group in
(\ref{hrju})). The main idea is to define the background field $x$
in (\ref{hrju}) in terms of the Polyakov loop $<TrW>$ with quantum
fluctuations included. This "classical" or "measured" value, $x_{cl}$,
can be calculated via $<TrW>$ and in the one-loop approximation the relation
between $x$ and $x_{cl}$ has the form
\begin{equation}
x=x_{cl}+g^2f(x_{cl}) +o(g^2)
\label{92}
\end{equation}
\noindent
where $f(x_{cl})$ is a function which has to be found. After the redefinition
the effective action reads
\begin{equation}
W(x)=W(x_{cl})=W^{(1)}(x_{cl}) + g^2[\frac{d}{dx_{cl}}W^{(1)}(x_{cl})
f(x_{cl}) + W^{(2)}(x_{cl})] + o(g^2).
\label{93}
\end{equation}
\noindent
It has been expected that at least the minimum position in $W(x_{cl})$ is
gauge invariant. After calculating the Polyakov loop in the one-loop
approximation the following relation has been obtained:
\begin{equation}
x=x_{cl}+\frac{g^2}{4\pi^2}B_1(x_{cl}/2)(\xi - \xi_0)
\label{94}
\end{equation}
\noindent
where $\xi_0$ is an arbitrary fixed number and $x=x_{cl}$ when $\xi=\xi_0$.
It is the gauge where the renormalization of the gluon propagator is absent.
Substituting (\ref{94}) into the effective potential and taking into
account the explicit form of the Bernoulli polynomials, the final result
for the two-loop effective action has been found
\begin{equation}
\beta^4W(x_{cl})=\pi^2[-\frac{1}{15}+\frac{1}{12}x_{cl}^2(x_{cl}-2)^2]
+ g^2[\frac{1}{24}-\frac{5}{96}x_{cl}^2(x_{cl}-2)^2]+o(g^2).
\label{95}
\end{equation}
\noindent
This functional has a minimum at $x_{cl} = 0$, which means the absence of the
condensate at two-loop level.  Now, we analyze this procedure in more
details. Actually, it interpolates between Green's function methods (described
above) and completely gauge invariant methods like the lattice formulation
discussed in the previous chapter. Really, in the Refs. \cite{belyev3},
\cite{bhat} as a first step the Green functions are used to calculate the
effective action. As a second one the reparametrization of the background field
in terms of $<TrW>$ is performed. The latter may occur to be not consistent.
The point is that in the loop expansion method, as a rule,
the background field  $A_0$ is chosen to be a solution of
classical field equations and is to be considered as a fixed parameter
through all calculations. Its vacuum value must be determined
via the minimization procedure. On the other hand, if one expresses $A_0$ in
terms of colourless parameter $<W>$ (but, in principle, other parametrizations
are possible) with quantum corrections to be taken into account and substitutes
it as an argument in the effective action then the variation of $W(A_0)$  with
respect to $<W>$ may not correspond to the determination of vacuum value
$<A_0>$ considered as the dominant classical configuration in the initial
partition function. It seems to us that it would be more natural to express
$W(A_0)$ in terms of $<W>$ from the very beginning and apply a loop
expansion to this functional. This scheme was described in the
previous chapter and we shown that the condensate does appear in such
a treatment.

After these remarks it will be very instructive to describe Belyaev's result
in terms of variables on characteristics. The basic point here is that the
relation (\ref{94}) and "observable" background fields coincide identically
with the characteristic in the $(x,\xi)$-plane which passes through an
arbitrary
point $(x_0 = x_{cl},\xi_0)$. In order to find the value of $W(x,\xi)$ on the
characteristic one must substitute  eq.(\ref{94}) in eq.(\ref{hrju})
(for the $I$-spin subgroup only in $SU(2)$ case) and then expand in powers of
$g^2$ to order $g^2$. After that one obtains the equation
\begin{eqnarray}
\beta^4W_{charact}(x_0,\xi)=\frac{2\pi^2}{3}[B_4(0)+2B_4(x_0/2)] +
\frac{g^2}{2}[B^2_2(x_0/2)+2B_2(0)B_2(x_0/2)]	\nonumber   \\
+ \frac{2}{3}g^2(1-\xi_0)B_3(x_0/2)B_1(x_0/2)
\label{96}
\end{eqnarray}
\noindent
As has been expected, on the orbit the effective action is independent on the
parameter $\xi$ but it is a constant depending on $x_0,\xi_0$. The minimum
position and minimum value of $W_{charact}$ are
\begin{eqnarray}
(x_0(\xi_0))_{min}=\frac{g^2}{8\pi^2}(3-\xi_0), \\
\beta^4W_{charact}((x_0)_{min},\xi_0)=-\frac{\pi^2}{15} +
\frac{g^2}{12} - \frac{g^4}{192 \pi^2} (3-\xi_0)^2.
\label{97}
\end{eqnarray}
\noindent
Taking into account that $(x_0)_{min}$ can be identified with $x_{cl}$ one
comes to the conclusion that $(x_0)_{min}$ in eq.(\ref{97})
is the gauge invariant "measured" value of the gluon condensate.
This is the exact meaning of the $\xi$-independence of $W_{charact}(x_0,\xi)$.
Moreover, in this way the idea to
express $W(A_0,\xi)$  in terms of the Polyakov loop can be realized in the
Nielsen identity approach. As we remarked before,
this possibility appears due to
the fact that characteristics in the $(x,\xi )$-plane and the
relation of $x_{cl}$ and $x$ are given by the same equation (\ref{94}).
If one puts $\xi_0=3$ in eqs.(\ref{96}), (\ref{97}),
the effective action (\ref{95}) and other results of Refs.
\cite{belyev3}, \cite{bhat} immediately follow. Hence it is possible to
conclude that Belyaev's method corresponds to
the choice of the orbit which passes through the point
$x_{cl}=x_0=0, \ \xi_0=3$.
This is a special gauge where $x_{cl} \neq 0$ might be determined
in the three-loop approximation. All other gauges signal $<A_0> \neq 0$ at
two-loop level.

In Ref. \cite{kalnew}, \cite{kalfian} the gauge-invariant thermodynamical
potential $\Omega (A_0)$ has been calculated step by step in perturbation
theory. In this method the minimum position of the next order of the potential
$\Omega^{(n)} (A_0)$ has to be calculated via the effective action
$W^{(n-1)}(A_0,\xi)$ of the previous order. In this way the result $<A_0>=0$
in two-loop approximation has also been obtained. But at the same time the
non-trivial minimum position for the three-loop thermodynamical potential
$\Omega^{(3)} (A_0)$ was found. Hence, a good chance for determination of the
gluon condensation at this level has appeared in the gauge-invariant method of
calculation. Thus, we come to the idea that a final determination of
$A_0$-condensate needs a calculation of higher loop contributions.
This is not a
surprise because $<A_0> \neq 0$ is the two-loop effect and to prove it
the first quantum corrections should be calculated.
This very complicated mathematical
task has not been solved yet. At the present time in the literature only some
partial results of the role of the higher loops have been reported. For
completeness we shall describe them here.

Generally speaking, when the Green function method is used all the observables
and, in particular, the higher loop contributions should be calculated with
Nielsen's identities taken into account. In calculation of
$W(A_0,\xi)$ it may occur that some diagrams of special kind give
$\xi$-dependent contributions. Then the only way to check the correctness
of the approximation scheme is to apply the Nielsen identities.
 From this stand point it is very essential,
as we have been convinced before,  that the loop expansion
is a consistent approximation scheme and no other $\xi$-dependent diagrams
should be added to $W^{(2)}(A_0,\xi)$  in order $g^2$. Only $\xi$-independent
contributions can be included. Keeping these arguments in mind,
let us discuss the results of Refs.\cite{belyev2},\cite{saw} where
the ring diagrams
(plasmon diagrams) $W_D(A_0)$ have been calculated in the Feynman gauge
$(\xi =1)$. As the main result it has been found that when one considers the
sum
of $W_D$ and $W^{(2)}$ the solution $<A_0>=0$ in the order $g^2$ follows.
However, from the above remarks it is clear that the conclusion
contradicts to what Nielsen's identities tell us.
To understand the situation in details we have
calculated the contribution of the ring diagrams $W_D(A_0, \xi )$ in the
$R_{\xi}^{ext}$ background gauge. It was found that this contribution is
$\xi$-dependent as well as $W^{(2)}(A_0, \xi )$.
Moreover, if we consider the sum
$W^{(2)}(A_0, \xi ) + W_D(A_0, \xi )$ the result $<A_0>=0$ follows again,
as in \cite{belyev2}. At the same time, when one substitutes $W_D(A_0, \xi )$
in the identity (\ref{91}), it does not fulfilled. So, the contribution
$W_D(A_0, \xi )$ is inconsistent. The resolution of this contradiction lies
in the way of calculation of $W_D(A_0, \xi )$ in \cite{belyev2},\cite{saw}.

As is known \cite{5kal}, in QCD and QED (with $A_0=0$) the sum of the ring
diagrams describes the contribution of infrared divergencies to the effective
action and results in the non-analytic term of order $g^3$. In this case the
infrared limit is calculated as follows: $k_{n=0}=0, \mid \vec{k} \mid
\rightarrow 0$ \cite{5kal}. The same definition is used in \cite{belyev2}
in the case of $A_0 \neq 0$. But actually this may not be the case and
an other definition must be considered.
In \cite{kalfian} the following one has been
introduced: $k_0=k_4+g\beta A_0=0, \mid \vec{k} \mid \rightarrow 0$.
Besides, the additional contribution resulting from the transversal part of the
polarization tensor has also been  calculated \cite{kalfian}. After that the
obtained corrections appear to be gauge independent  and in the limit
$A_0 \rightarrow 0$ reproduce the result for the $A_0=0$ case. So, just
this non-local term should be included to the effective action
$W(A_0, \xi )$. With this contribution we have no $A_0$-condensate elimination.

Other essential result for understanding the properties of the higher loop
effects has been reported in  \cite{kalnew} where, in particular,
the position of the three-loop thermodynamical potential $\Omega^{(3)}(A_0)$
was determined which is both non-trivial and $\xi$-dependent. Hence,
the hope to determine $<A_0> \neq 0$ at the three-loop level has obtained a
real
confirmation. Anyway, the calculation of the three-loop contribution to the
effective action remains of great importance for the final solution
of the discussed problem.

\section{Effective Lagrangian approach to $A_0$ condensation}

It is well-known that gauge fields are invariant under global
$Z(N)$-transformations. But at finite temperature the Polyakov loop
transforms under $Z(N)_{gl}$ subgroup as matter fields in the fundamental
representation (\ref {19}). This implies that eigenvalues of the Polyakov
loop or $A_0$ have the charges of centre of $SU(N)$ like quarks.
Besides, the Polyakov loop transforms under $SU(N)_{gl}$ group like the
matter fields in the adjoint representation (\ref {27}). It is a
motivation to consider $A_0$ as the Higgs field. Basically, the action
$S^G$ (\ref {25}) has the Higgs form for $\varphi_x$ defined in
eq.(\ref {26}). These fields are living in the lattice sites, they have
non-trivial self-interaction and can provide a minimum far away from
zero. So the action $S^G$ is more likely to be the "standard" $SU(2)$ Higgs
lattice action for the $\varphi$-fields.
The evidences that at high temperature $A_0$ indeed behaves as the Higgs
field in the continuum theory and its Lagrangian has the Higgs form
have been given in\cite {polvaz,bogacek,olesz}.

In the continuum theory the minimum of the pure gauge action is reached
for the semiclassical values of the potentials $A_0^a = const.$ and
$A_i^a = 0.$ If we admit a semiclassical value of $A_0^a \neq 0$
we immediately find the massive chromomagnetic gluons in the effective theory.
To obtain the action for the vacuum corresponding to $A_0^a \neq 0$
we integrate out these massive modes following the philosophy of
Appelquist-Carrazone decoupling theorem \cite{app1}. As a result we obtain
an effective Higgs potential for quantum fluctuations of field $A_0$
treated as the Higgs field. This Higgs potential is periodic as a
function of chromoelectric gauge field with period $\frac{2\pi}{g\beta}$.
Due to this periodicity the new
vacuum of the effective  system possesses the periodic symmetry
$$
A_0^a \rightarrow A_0^{a} = A_0^a + \frac{2 \pi}{g \beta}
$$
\noindent
together with the parity symmetry
$$
A_0^a \rightarrow A_0^{a} = -A_0^a.
$$

For $A_0$ within the interval of periodicity the effective $SU(2)$ three -
dimensional theory has been built and it has been shown that the global
gauge symmetry is broken due to formation of non-zero expectation value
of $A_0$. On the contrary, in the case where all the semiclassical potentials
vanish the corresponding effective system is completely different.
We find there massless gluons interacting with Higgs field in a state
with unbroken symmetry characterized by the effective
Higgs potential with global minimum for $A_0=0$ displaying only
parity symmetry. Transition to the
system described earlier must be accompanied by a symmetry breaking process.

Let us now give a sketch of the calculation and  show how the global
gauge symmetry breaking happens. The Fourier expansion with respect to the
compactified dimension enables us to reduce the
four dimensional theory to the effective three dimensional field model.
The Fourier field components acquire frequency dependent masses
and therefore it looks reasonable to apply for nonzero
frequency modes the Appelquist-Carazzone decoupling theorem\cite{app1}
to calculate an effective theory for static field modes. This process
generally known as dimensional reduction was described by Appelquist and
Pisarski\cite{app2} as a possible calculation scheme for high temperature
behaviour of field theories. Following this approach, an effective
static model in three dimensions
was proposed, reasonably describing the full four dimensional
theory in the distance range $RT \cong 1$ at sufficiently high
temperature in the deconfinement phase\cite{bogacek}.
The idea of calculation consists in a reduction of
a number of degrees of freedom in the continual integral definition
of the partition function and/or the thermal mean values of the observables.
The Fourier expansion of the fields with respect to the compactified direction
was applied which enables us to integrate out the
nonstatic modes of the fields. The frequency dependent ``mass'' term
prevents the procedure from infrared divergencies. The ultraviolet
divergencies are regularized by zeta function regularization method.
This procedure results in a nonlocal functional
determinant. Expanding the determinant we obtain finally
the effective action for static modes in three dimensions. This
expansion is obtained within the background field method for
the chromoelectric field in the static gauge
\begin{equation}
A^{3}_{0}({\bf x}) = A_{0}^{B} + \varepsilon({\bf x})
\label{98}
\end{equation}
\noindent
where $\varepsilon({\bf x})$ represents the quantum field fluctuations
around the constant field value $A_0$. In the effective theory the
chromoelectric field is identified with the  Higgs field in the
adjoint representation\cite{polonyi}. The effective
action depends explicitly on the static field value $A_{0}$ through
the effective Higgs potential. The mean values of operators
calculated by this method must be complemented by the information
about the value of $A_{0}$ in the minimum of the effective potential.
Depending on the number of degrees of freedom chosen we can calculate
the different effective systems. To decide which effective system
for fixed finite temperature corresponds to the full system is the question
of comparison the minima of the effective potentials corresponding to the
different effective actions derived from the action of the full system.
We suppose that in our approach the partition function is not changed by
the mathematical operations in the process of the calculation of the
different effective systems.
Therefore the system with lower minimum of the effective potential
is the most probable  effective system corresponding to the full theory
at fixed temperature.

Starting from $SU(2)$ gluodynamics we have used the action
(\ref{15}) expanded up to the second order in the chromomagnetic
fields. Following the calculation described above we come to the effective
potential for $A_0^B$ of the form:
\begin{equation}
U_{eff}(X)=-\beta^{-4}(\frac{6\pi^2}{90}-\frac{4\pi^2}{3}[X^4-2X^3+X^2])
-\frac{1}{V\beta} \ln Det^{-\frac{1}{2}} (\frac{S_{cor}(\chi
> \chi_{min}, X)}{\chi^2})
\label{99}
\end{equation}
\noindent
where
\begin{eqnarray}
\lefteqn{S_{cor}(k,X) =\frac{(4\pi l)^2}{2}} \nonumber \\
&& + g_r^2 \left\{
2(X^2-X+\frac{1}{6}) - \frac{11}{3}l^2[(1-\ln(4\pi )-\Psi (X)-\Psi(1-X)]
+\frac{l^3\pi^3}{\sin^2(\pi X)}  \right.  \nonumber \\
&& + \sum_{-\infty}^{\infty} \left[-\left(
\mid n+X\mid^2 + \frac{2l^4}{\mid n+X\mid^2} +2l^2 \right)
\frac{1}{l}\arctan\frac{l}{\mid n+X\mid}  \right. \\
&& \left. \left. + \mid n+X\mid + \frac{5}{3}\frac{l^2}{\mid n+X\mid }
\right] \right\} \nonumber
\label{100}
\end{eqnarray}
\noindent
Here we have used the new field variable $X=g \beta A_0/(2\pi)^{-1}$,
$g_r$ is the temperature dependent running coupling constant,
and $4\pi{\bf l}=\beta{\bf k}$ is the dimensionless momentum.
The functional determinant
in (\ref{99}) is calculated by integration over variable $\chi$
by means of the $\zeta$-function regularization scheme.
Analyzing the last equation we find the following picture.
For larger coupling constant (low temperature) the extreme
of the $U_{eff}(X)$ is achieved at $X=\frac{1}{2}$. This value corresponds
to the confinement phase where the expectation value of the Polyakov loop is
equal to zero. When coupling constant is decreasing the minimum value of the
potential is removing from the point $X=\frac{1}{2}$  and there will appear
two symmetric minima which signal that $SU(2)$ global gauge symmetry is
broken as in these minima $X \neq n$ where $n$ is any integer.
This means that in this phase the condensate falls,
in full accordance with our previous investigations.
In the same time we found that the Polyakov loop
in this phase differs from zero. In this sense this calculations are very
similar to those presented in chapter 4 where the $<A_0> \neq 0$ generation is
accompanied by deconfinement phase transition. Our approach differs from that
of Weiss one \cite{anish} by taking into account the quantum fluctuations
around $A_0^B$ gauge potential.

It is  interesting to compare the calculations presented above with
the similar ones done by Oleszczuk in  \cite{olesz} where it has been shown
that $\langle A_0\rangle \neq 0$ does not appear at one-loop level.
We find the differences in treating both the chromoelectric part of the
initial action and massive static modes of the chromomagnetic potentials.
In our approach the massive static modes of the chromomagnetic potentials
were included in the effective potential whereas in \cite{olesz} these modes
were missed. As the result the Oleszczuk vacuum state contains two states
with mass term proportional to $(A_0^B)^2$. It implies that the effective
potential "wants to correct " the situation and final expression for the
$U_{eff}$ does not contain the third order of the background field and
possesses the minimum for $A_0^B =0$. In this minimum the masses proportional
to $(A_0^B)^2$ vanish and the former massive modes become massless
and, therefore, they are a part of the vacuum.

Nevertheless, if we build the semi-classical expansion around the gas of
chromomagnetic monopoles\cite {olesz} the result $\langle A_0\rangle \neq 0$
can be recovered.

As we promised in chapter 4, we are now discussing the problem of the
Vandermonde determinant cancellation. In the static gauge this
determinant appears as the Fadeev-Popov determinant and
can be calculated directly without the ghost formalism.
In \cite{bogacek} it has been shown that the Fadeev-Popov
determinant is indeed cancelled by contributions coming from the integration
over transversal chromomagnetic potentials. It happens, as was discussed
in \cite{pisar}, even beyond one-loop level. However, it does not lead to the
conclusion that $A_0^B=0$ in the vacuum \cite{pisar}. In fact the determinant
is cancelled from the effective potential even when $A_0^B \neq 0$
\cite{bogacek}. This fact has very simple and beautiful explanation.
As we pointed out earlier the integration over chromomagnetic gauge field
in the present approach is equivalent to projection onto gauge-invariant
states. It means that if the cancellation takes place when $A_0^B=0$
the same must happen also for $A_0^B \neq 0$ if the latter is a
gauge-invariant phenomenon.
 From this fact we can deduce that the present calculations
performed in the static gauge lead to the gauge-invariant results since
the mentioned cancellation does take place.

\section{$A_0$-condensate in theory with dynamical quarks}

As it follows from the title, this chapter will be devoted to studying
the influence of the dynamical quarks on the condensate.
We may say that quarks do not change a general picture in essential
way and both lattice strong coupling approximation with massive
quarks \cite {bor}, \cite{bogolyubov} and loop expansion with massless
quarks \cite {scl8}, \cite{chub} have shown that condensate does not
disappear. Nevertheless there are some disagreements between these
two approaches which should be clarified. Let us begin with calculation
of the quark contribution on the lattice.

Two approaches can be developed to solve this task. In \cite{bogolyubov}
the hopping parameter expansion was utilized to include the effects of
massive quarks in Euclidean version of the lattice theory. In \cite{bor}
the quark contribution was obtained in Hamiltonian formulation by means of
Banks-Ukawa method \cite{banks} and calculation of first not-trivial term
of high temperature expansion of the fermionic determinant. Since the results
have been obtained to be essentially the same, we limit ourselves here to
elaborate on the second approach as it looks as being reliable in more
broader range of parameters.

Lattice Hamiltonian for Kogut-Susskind fermions is of form
\begin {eqnarray}
S_{K-S} = \frac{1}{2}\sum_{x,n=-d}^{d} \eta_{n}(x)\bar{\Psi}(x)
U_{n}(x)\Psi(x+n) + m_qa \sum_{x}\bar{\Psi}(x)\Psi(x), \nonumber    \\
\eta_{-n} = -\eta_{n}, \  U_{-n}(x)=U_{n}^{+}(x-n)
\label{101}
\end {eqnarray}
\noindent
where $\eta_{n}(x) = (-1)^{x_{1}+x_{2}+...+x_{n-1}}$.
We remind that we should now use the projection operator (\ref{49})
where
\begin{equation}
q^a = \bar{\Psi}(x) \lambda^a \Psi(x).
\label{102}
\end{equation}
\noindent
The integration over the quark part of the Hamiltonian yields the factor
$Z_q$ and for the massless quarks (for the sake of simplicity let us
take it for a short while) the result turns out to read
\begin{equation}
Z_q=Z_q^0 det(I+\frac{\delta_{ab}}{q_x^a}\frac{\beta}{a_{\sigma}}D_{xy}^{ab})
\label{103}
\end{equation}
\noindent
where we have noted
\begin{equation}
q_x^a = 1 + \exp (i\phi_x^a)
\label{104}
\end{equation}
\noindent
and
\begin{equation}
D_{xy}=\frac{1}{2}\sum_{n}(U_{xy}\delta_{y,x+n} -U_{xy}^+ \delta_{y,x-n}).
\label{105}
\end{equation}
\noindent
The diagonal part of the fermionic determinant $Z_q^0$ resulting from
the projection operator is explicitly given as \cite{banks}
\begin{equation}
Z_q^0=\prod_{x}[1+ Re \Omega (\phi_x)].
\label{106}
\end{equation}
\noindent
For non-diagonal part of the determinant the high-temperature expansion can
be constructed. The necessary details can be found in \cite{borpet}.
After calculation we come to the following quark contribution restricting
ourself to the first non-trivial term in high temperature expansion
\begin{equation}
-S_{eff}^{Q}= \ln Z_q^0+ \ln (\int \prod_{x,n}dU_n(x)
\exp [-\frac{\beta^2}{8a^2_{\sigma}}Tr \sum_{x,n}\frac{1}{q}U
\frac{1}{q^{\star}}U^{+} ....]).
\label{107}
\end{equation}
\noindent
The dots represent here contributions of the next even terms of logarithmic
function expansion. All these terms are negative therefore we can regard
the maximal values of $Z_q^0$ to correspond to the minimal values of the
quark effective potential. It is clear that maximum of $Z_q^0$ is achieved
when $<\phi>=0$. This result is very essential as it shows the
gluon condensate generation is caused by gluonic kinetic energy only but
not  the quark contributions. This point is to be in disagreement with
the loop expansion method where condensate appears either from gluon or from
quark sector (see discussion below).

Further we choose to work with $SU(3)$ gauge theory.
As in the pure gluodynamics, there are six solutions here, however, due to
the explicitly broken $Z(N)$ symmetry they are not equivalent. Now the basic
maximum is developing at $<\phi_1>=<\phi_2>=0$. It is clear that evaluating
the effect of massive quarks a la Banks-Ukawa \cite{banks} i.e. making
the substitution
\begin{equation}
q_x^a \rightarrow  \exp (\beta m_q) + \exp (i\phi_x^a)
\label{108}
\end{equation}
\noindent
we come to the same conclusion: condensate does not appear at any
temperature.

Performing now the invariant integration in (\ref{107}) we have up
to an irrelevant constant
\begin{equation}
-S_{eff}^{Q}= \ln [1+ Re \Omega (\phi)] - \frac{d \beta^2}{16a^2_{\sigma}}
\frac{[Im \Omega (\phi)]^2}{[1+ Re \Omega (\phi)]^2}
\label{109}
\end{equation}
\noindent
and then combining (\ref{60}) and (\ref{109}) with substitution
(\ref{108}) we have analyzed the effective potential
$S_{effG} + S_{eff}^{Q}$ at different values of $\gamma$ and $m_q$ and found
the following picture. At larger $\gamma$ there is one minimum
where condensate is absent. With temperature increasing ($\gamma$ decreasing)
a phase transition takes place at $\gamma_c^q$ which is somewhat larger
than for pure gluodynamics. Below $\gamma_c^q$ the effective action
develops  two minima where $<\phi_1> \neq 0, <\phi_2> \neq 0$.
It implies that besides spontaneous breaking of the global gauge symmetry
the condensate generation may lead to spontaneous breaking of the colour
charge symmetry. This conclusion is in some contradiction to the loop
expansion method as well. Such a breaking means that in the corresponding
minima the baryonic number is generated. We elaborate this result in details
in the next chapter. However, at very high temperature when quark mass
is vanishing we have found that the deepest minimum of the effective action
appears to be $C$-symmetric again with $<\phi_1> \neq 0, <\phi_2> = 0$.
In such a way the disagreement with loop calculations can be avoided.
Hence, our more essential conclusion is that $A_0$-condensate falls at high
temperature in full QCD. $<A_0>$ is non-vanishing because of fluctuations
of the gluonic kinetic energy and does not present in the quark sector
of QCD.

Now let us present the results of loop calculations in the continuum QCD with
dynamical quarks. All our notations follow chapter 5 where we discussed
the loop approach in gluodynamics. The one-loop contribution of
massless quarks to the $W(A_0)$ has been calculated in \cite{gross}. The
two-loop quark functional was obtained in Refs.\cite{chub}, \cite{sclnew}
The Feynman rules and necessary integrals and sums are listed in
Appendix A of Ref. \cite{chub}.
Thus, up to two-loop order we have
\begin{eqnarray}
    &&W_{q}(A_0)\frac{\beta^4}{N_{f}} =
    -\frac{4\pi^2}{3} \sum^{3}_{i=1}\left\{
    B_4(c_i)\right\}		   \nonumber \\ &&-\frac{1}{2}g^2
    \left\{\right.  B_2(a_1)\left[B_2(c_1)+B_2(c_2)\right]
    +B_2(a_2)\left[B_2(c_1)+B_2(c_3)\right]		    \nonumber \\
    &&+B_2(a_3)\left[B_2(c_2)+B_2(c_3)\right]
    -\frac{1}{3}\sum^{3}_{i=1}\left[B_2^2(c_i)-2B_2(0)B_2(c_i)\right]
							    \nonumber \\
    &&-B_2(c_1)B_2(c_2) -B_2(c_2)B_2(c_3)
     -B_2(c_3)B_2(c_1)	\left.\right\}			     \nonumber \\
    &&+\frac{(\xi-1)}{3}g^2
    \left\{\right.
     ~~B_1(a_1)\left[B_3(c_1)-B_3(c_2)\right]			       \\
    &&+B_1(a_2)\left[B_3(c_1)-B_3(c_3)\right]
    +B_1(a_3)\left[B_3(c_2)-B_3(c_3)\right]
    \left.\right\},					     \nonumber
\label{110}
\end{eqnarray}
\noindent
where $x,y$ and $a_i$ are defined in (\ref{67}) and
\begin{eqnarray}
    &&c_1=\frac{1}{4}(x+\frac{1}{\sqrt{3}}y+2),~~~~~~
      c_2=\frac{1}{4}(-x+\frac{1}{\sqrt{3}}y+2),       \nonumber   \\
    &&c_3=-\frac{1}{2\sqrt{3}}y+\frac{1}{2}~~~~~~~~~~~
\label{111}
\end{eqnarray}
\noindent
where $N_f$ is a number of quark flavors, $B_{i}(x)$ is the Bernoulli
polynomial
of the order $i$ defined modulo 1. Calculating the minimum position and minimum
value of this effective action we obtain
\begin{eqnarray}
\beta^{4}W_q=-\pi^2\frac{7}{60}N_f + g^2\frac{5}{72}N_f
		    -g^{4}N_{f}\frac{(3-\xi)^2}{192\pi^2},     \nonumber \\
   x_{min}=g^{2}\frac{3-\xi}{4\pi^2},~~ y_{min}=0.
\label{112}
\end{eqnarray}
\noindent
Here, the minimum values were calculated for  intervals
\begin{eqnarray}
    0\leq a_1\leq 1,~ 0\leq a_2\leq 1,~ -1\leq a_3\leq 0, \nonumber  \\
    0\leq c_1\leq 1,~ 0\leq c_2\leq 1,~  0\leq c_3\leq 1
\label{113}
\end{eqnarray}
\noindent
and, as in the pure gluodynamics, five other minima in $(x,y)$-plane
can be found by means of consequent rotations by the angle $\frac{\pi}{3}$.
It is interesting to notice that the value of the condensate is the same
both in the gluon and in the quark contributions. Moreover, both  fermions
and bosons act to lower the action of $A_0$ field. This is rather unusual
property that the condensate is produced in both sectors of the theory
separately. As a total result, we have in the two loop approximation
\begin{equation}
\beta^4(W_q+W_{gl})_{min}=-\pi^2(\frac{8}{45} + \frac{7}{60}N_f) +
g^2(\frac{1}{6} + \frac{5}{72}N_f) -	      \nonumber    \\
g^{4}\frac{(3-\xi)^2}{32\pi^2}(1+\frac{N_f}{6})
\label{114}
\end{equation}
\noindent
It is worth mentioning that $y_{min}=0$ in the deepest minimum
(what means - in the vacuum). Hence, it follows that $A_0$-condensate does
not effect a baryon charge at high temperature. Now let us elaborate the
vacuum structure of gluon condensate which results from (\ref{68}) and
(\ref{114}). The quarks change the symmetry of the vacuum. As is well
known, the Bernoulli polynomials are defined modulo 1. Hence, the effective
action possesses the symmetry in the $(x,y)$-plane. We display this symmetry
in the Fig.1. Any dot in the plane can be translated along dashed lines.
Vacuum structure consists of the hexagonal and triangular elements,
which cannot be transformed one to another by translations. At the
translations the hexagonal elements pass to themselves.
So do the triangular ones.

{The global minima of the effective
potential  $W_q+W_{gl}$ are marked by dots.
Besides, in each triangular area there are six local minima which
are not depicted.  The local minima appear only if
the quark and gluon contributions are included together. They disappear
when one considers the gluon contribution only.
The local minima depths decrease and the global ones increase while the
number of quark flavors $N_f$ becomes bigger .}\\
At some $N_f$ the local minima disappear completely.
To understand  the effect of quark mass we also have calculated
the contribution (\ref{110}) with $m_q \neq 0$. Unfortunately, this calculation
is incomplete because of complexity of integrals and the result has been
obtained for small $m_q$ only. It qualitatively looks as follows.
The difference between the local and global minima decreases
with the quark masses increasing. The local minima are equal to global ones
and we obtain the same result as in the gluodynamics. Thus, mainly the light
quarks bring up to an appearance of local minima which	can be identified
with metastable phases of the quark-gluon matter.
So, one can assume that at some intermediate temperature the phase
transition from the global minimum to the local one takes place.
This transition may lead to $y_{min} \neq 0$ in a vacuum. So, it can be
expected that the chiral phase transition is accompanied by the baryon
number generation and by spontaneous breakdown of the charge symmetry.

Now let us check the identity (\ref{91})  for the quark contribution.
Differentiating the one-loop part of $W_q$ with respect to
$B^3=\overline{A}_0^3$ and $B^8=\overline{A}_0^8$ we have
\begin{eqnarray}
   \frac{\partial W^{(1)}_{q}}{\partial B^{3}_{0}}=
   -\frac{4\pi}{3 \beta^3}
   g[ B_{3}(c_1)-B_{3}(c_2) ],	 \nonumber   \\
   \frac{\partial W^{(1)}_{q}}{\partial B^{8}_{0}}=
   -\frac{4\pi}{3\sqrt{3}  \beta^3}
   g[ B_{3}(c_1)+B_{3}(c_2)-2B_{3}(c_{3}) ].
\label{115}
\end{eqnarray}
\noindent
Then by summing up the products of eq. (\ref{89}), (\ref{115})
the RHS of eq. (\ref{79}) can be calculated. The obtained expressions
coincide up to the sign  with derivative of (\ref{110}) with respect
to $\xi$. Thus the Nielsen identity (\ref{91}) is fulfilled
for quarks as in the case of gluons. Taking into account that non-vanishing
$A_0$-condensate lowers the actions $W_q, W_{gl}$ and $W_q+W_{gl}$ and the
above result we see that both basic properties	required by the Nielsen
identities approach are held. In accordance with general theory we
come to the conclusion that the gluon condensation at finite temperature
is a gauge invariant phenomenon with quarks included in consideration.

Thus, both lattice and loop expansion considerations led us to the similar
conclusion about $A_0$-condensate in the full QCD though the effect
of the quark contribution to the effective action has been found out
to be essentially different. This discrepancy does not have an
explanation at the moment
and should be clarified in the future investigations. The reason can be
the following. In fact, the high temperature expansion used on the lattice to
calculate a quark contribution is well reliable only at
large $m_q$ whereas the loop calculations have been performed at
vanishing quark masses. So it is very desirable to carry out both
calculations at intermediate values of quark masses. Just this point
can be the origin of another disagreement concerning a possibility of the
baryon number generation which will be discussed in the next chapter.

\section{$A_0$-condensate in hot gauge theories. Some consequences}

What would be the consequences of such a condensate? We shall present
some of them and give a brief review of the most important results
and related problems. We begin our discussion with
the problem which was actually one of that difficulties of hot
non-abelian gauge theories which has impelled to develop the approach
to these theories on the basis of the possibility of the global gauge
symmetry spontaneous breaking. This is infrared problem
which have been already discussed in different aspects.
Now, the question is: what may we call a solution of the infrared problem?
It has been shown in the first papers
by Linde that appearing of infrared cut-off even of the order $T$
(the condensate appears to be just of this order) cannot save the
perturbative expansion (in this case all terms of the perturbative expansion
starting from the order $g^6$ are proportional just to this order and
therefore all of them give equal in coupling constant contributions).
 From the mathematical point of view the answer is quiet clear -
we need to construct a method of calculation which would give  finite
results for expectation values of physical observables in the
continuum. One of such	methods we discussed in chapters 2 and 6.
This is the reduction of gauge theories at high temperatures
to the effective three-dimensional models.
It is possible to show that in the course of the reduction the static modes
of the chromomagnetic potentials acquire mass due to formation of the
$A_0$-condensate \cite{3nad2}, \cite{olesz}. Thus, in this method of
calculations a hope to avoid the infrared divergences by means of
the condensate does appear.
We think that the most	essential point
in proofing that $A_0$-condensate can cure the theory of the divergences is
the proof that just this condensate lead to screening all sources
in the adjoint representation at high temperatures. The existence of such a
screening  would mean, in our opinion, the solution of the infrared problem.
We remind here that we should consider spatial correlators to calculate
this screening (see our discussion in the introduction). Now, we are going
to show, omitting all technical details, how this screening can appear
in the reduction of lattice gauge theories \cite{borpet}, \cite{ildgt}.

We present a method much analogous in its idea to the recent development
of perturbative expansion resumming \cite{pisbr} conjectured to cure
the infrared divergences of finite temperature QCD.
Actually, the reduction looks like isolation of static contribution
in the action after Fourier transforming gauge fields $A_{\mu}(x)$
with further calculation of multi-loop corrections
over massive non-static modes. This procedure has been already
performed both in the continuum theory for the Yang-Mills action
(see previous chapter) and on the lattice for the Wilson action \cite{reis}.
In the latter the reduction  beyond the perturbative horizon
\cite{cock}, \cite{ris} can be accomplished as the gauge matrix
$U = \exp (gaA)$ expansion is not necessary.

We develop proper high-temperature expansion and the corresponding reduction
dealing with the Fourier transformation of compact gauge matrices
$U_{\mu}(x)$ rather than gauge fields $A_{\mu}(x)$ calculating then a static
contribution resulting from such an expansion.
Two essential points distinguishing this approach  from the preceeding
ones \cite{app2}, \cite{olesz}, \cite{reis} appear to be the following.
Firstly, the static sector generates compact dielectric field
leading to an effective dielectric theory. Secondly, since in
the present case Fourier transformation is not simple linear substitution
the Jacobian of this transformation is non-trivial and moreover it generates a
mass of dielectric field (which is, in fact, the mass of static modes).
As to the non-static modes they are massive as before \cite{app1}
with the mass proportional to $(nT)$ (in continuum limit).
Therefore, in this way we could construct quite reliable
perturbative expansion for massive modes.
All calculations of the resulting effective action can be found
in \cite{borpet}, \cite{ildgt} where both corrections to the main static
contribution and quark contribution have been computed. Here, we write
down the effective action for static modes restricting to the pure
gluodynamics.

\begin{eqnarray}
Z = \int \prod_{x}d \mu( \alpha_{x}) \prod_{x,n}\exp(\beta^{\prime}
\cos\varphi_{x} \cos\varphi_{x+n} - 2\lambda_{e}
\cos\alpha_{x} \cos\alpha_{x+n}) Z_{D},
\label{116}
\end{eqnarray}
\noindent
\begin{equation}
Z_{D} = \int \prod_{l}[\rho^{3}_{l}d\rho_{l} d\mu(U_{l})]\exp{S_{D}}
\label{117}
\end{equation}
\noindent
$\beta^{\prime} = 4(\lambda_{e}/ \lambda^{0})^{N_{t}}$,
$l = (x,n)$ is a link.
This effective action we call induced dielectric action.
In mentioned approximations we have
\begin{equation}
S_{D} = S_{E}^{stat} + S_{m}^{stat} + S_{mass}^{0},
\label{118}
\end{equation}
\noindent
where we have laid
\begin{equation}
S_{m}^{stat} = \lambda_{m}N_{t} Sp \sum_{\overline{p}}
\rho(\partial\overline{p}) U(\partial\overline{p}),
\label{119}
\end{equation}
\noindent
\begin{equation}
S_{mass}^{0} = - \frac{N_{t} \lambda^{0}}{2} \sum_{x,n}
Sp(\Phi_{n}(x)\Phi_{n}^{+}(x)),
\label{120}
\end{equation}
\noindent
\begin{equation}
S_{E}^{stat} = N_{t} \lambda_{e}\sum_{x,n}
Sp (\Phi_{n}(x)V_{x}\Phi_{n}^{+}(x)V_{x+n}^{\star}).
\label{121}
\end{equation}
\noindent
$V_{x}$ is zeroth component of gauge field matrices in the
static gauge and $\lambda_m = \frac{2a_{\sigma}}{g^2a_{\beta}}$,
$\lambda_e = \frac{2a_{\beta}}{g^2a_{\sigma}}$. In chosen
approximation $\lambda^0 \approx 2$. The representation
$\Phi_n(x)=\rho_n(x) U_n(x)$ was using in the course of the reduction.

Comparing formulae (\ref{117})-(\ref{121}) with (\ref{10})-(\ref{13})
from Introduction one can
easily deduce that (\ref{117}) is nothing  but some kind of LDGT.
There exist two differences from formulation \cite{mack3}
described in the introduction to be emphasized. First one is
that the theory (\ref{117}) is a compact since the dielectric field obeys
the relation $0 \leq \rho \leq 1$  unlike (\ref{12}).
Another significant difference
is the presence of the interaction term $S_{E}^{stat}$ which
describes interaction between gauge
field $U_n(x)$, dielectric field $\rho_n(x)$ and Higgs field $A_0$.

The mass of the static mode can be calculated from a "naive" effective
potential for dielectric field. Its classical form can be easily
obtained if we neglect the fluctuations of the gauge field
$U_{n}(x)$ and take the expectation value $<A_{0}>$
in $S_{D}^{H}$. The result has been put down in eq.(\ref{14}).
It is clear from its form that in the continuum limit the mass is
proportional to $<A_0>$ as we need to expand $\cos (a_{\beta}gA_0)$
when $a_{\beta}$ tends zero (temperature independent constant part
should be avoided from the mass after adding zero temperature contribution
to the free energy). Two facts following from the present consideration
should be stressed. Firstly, as was conjectured in Introduction,
two screening mechanisms - dielectric one and that caused by global gauge
symmetry breaking -  can be indeed related as $A_0$-condensate supplies
the dielectric field with gauge-invariant mass. Secondly, we have
proved that the mass of the static mode appears not only on the level
of the standard reduction but beyond the perturbative horizon if we
take into account all powers of static modes. Now, there is no
difficulty to prove that spatial adjoint Wilson loop obeys perimeter law at
any temperature \cite{ildgt}.
Moreover, it is possible to prove that the fundamental
Wilson loop displays the area law behaviour with string tension
to be proportional to the mass of the dielectric field \cite{ildgt}
in full accordance with results  of \cite{9bor}, \cite{10pol}.
All of that, according to duality relations,
mean the screening of the gluon currents and confinement of the
"spatial" sources being in the fundamental representation. Thus, we can see
that $A_0$-condensate does lead to the screening of the chromomagnetic forces
at high temperature. (Of course, our proof is not  strict and more exact
calculation of the adjoint Wilson loop in the effective theory with the
corrections from non-static modes included is very desirable.)

The next interesting problem we are going to discuss here is evaluating
the heavy quark potential in background $<A_0>$. As is known, heavy
quark potential at finite temperature is calculated in different phases
by using the correlator of the Polyakov loops.
We find the linearly rising potential in the confinement phase and
the Debye screened potential in the deconfinement phase where $<A_0> \neq 0$.

The heavy quark potential $V_{qq}$ in the finite temperature theory is
connected with the Polyakov loop correlator as \cite{yasve}
\begin{equation}
{\rm e}^{-\beta V_{qq}({\bf x_{1}},{\bf x_{2}})}=
\langle L({\bf x_{1}}),L({\bf x_{2}}) \rangle
\label{122}
\end{equation}
\noindent
In the static gauge we have for the Polyakov loop the simple
expression ($SU(2)$ gauge group)
\begin{equation}
L({\bf x}) = \cos (\frac{1}{2}\beta g (A_0 +\varepsilon ({\bf x})),
\label{123}
\end{equation}
\noindent
i.e. Polyakov loop depends on the chromoelectric field only. This justifies
us to use for the calculation of the correlator the same method as we
have used for the calculation of the effective potential in the chapter 6.
The effective potential was
obtained from the relation for the partition function expressed in terms
of the effective action \cite{bogacek}:
\begin{equation}
Z(\beta ) = \int [D\varepsilon ] {\rm e}^{S_{eff}}
\nonumber
\end{equation}
\noindent
For the effective action, applying the method described in chapter 6,
we have found the relation:
\begin{equation}
S_{eff}= \frac{V}{\beta ^3} \left[\frac{3\pi ^2}{45}-\frac{4\pi ^2}{3}
(X^4-2X^3+X^2)\right]-8\int d^3l\ \tilde{\varepsilon}({\bf l})
S^{eff}_{cor}(k,X) \tilde{\varepsilon}(-{\bf l}).
\end{equation}
\noindent
In the last expression $\tilde{\varepsilon}({\bf l})$ is the Fourier
transform of the function $\beta \varepsilon ({\bf x})$,
$S^{eff}_{cor}(k,X)$ has been put down in (\ref{100}) and all other
notations follow chapter 6.

By the same method one finds for the correlator of two
Polyakov loops at the distance $R$ the equation:
\begin{equation}
\langle L(0),L(R) \rangle = \frac{1}{2}{\rm e}^{-\frac{g^2_r}{8} S^{-1}(0,0)}
[{\rm e}^{\frac{g^2_r}{8} S^{-1}(0,R)} + \cos (2\pi X)
{\rm e}^{-\frac{g^2_r}{8} S^{-1}(0,R)}].
\label{124}
\end{equation}
\noindent
$S^{-1}$ means the inverse matrix element
with continuum indices given by the relation:
\begin{equation}
S^{-1}({\bf x},{\bf y}) = \int \frac{d^3 (\beta k)}{(2\pi )^3}
\frac{1}{S^{eff}_{cor}(k,X)} {\rm e}^{-i{\bf k}({\bf x} - {\bf y})}
\label{125}
\end{equation}

The direct application of last formula is not possible on the level of
the approximations defined in\cite{bogacek}, because $S^{eff}_{cor}$ is
negative
for small $k=\mid{\bf k}\mid$ in the neighborhood of $X = 1/2$. This is not
consistent with the definition of  physically acceptable action in the
functional integral. To remove this inconsistency in a natural way
we carefully review the terms abandoned in the previous method of
the calculation\cite{bogacek} with the hope to find some contributions making
$S^{eff}_{cor}$ positive. The calculation is in progress now and we
would like to stress its significance by pointing out
the important consequences.

Here, we simply modify $S^{eff}_{cor}$ to the positive function
$S^{mod}_{cor}$ by addition of a small positive constant. We rewrite
Eq. (\ref{125}) in the form:
\begin{equation}
S^{-1}(0,R) = \frac{1}{i(2\pi )^2\rho } \{ \int_{-\infty}^{0}
\frac{l\ dl}{S_{cor}^{mod}(-l,X)}\ {\rm e}^{il \rho}
+ \int_{0}^{\infty }
\frac{l\ dl}{S_{cor}^{mod}(l,X)}\ {\rm e}^{il \rho} \}
\label{126}
\end{equation}
\noindent
where $\rho = T R$.

The singular structure of the integrand in the complex $l$ plane is defined
by the singular structure of the $S^{mod}_{cor}$.
The function $S^{mod}_{cor}(l,X)$ possesses two cuts, $(-i\infty, -iX), \\
(iX, i\infty)$ in the complex $l$ plane.
Because the integrands in Eq.(\ref{126}) have no singularities
out of real and imaginary axes, moreover,
\begin{equation}
\lim_{\mid l \mid \rightarrow \infty} l \left[S_{cor}^{mod}(l ,X)
\right]^{-1} =0, \nonumber
\end{equation}
\noindent
we change the range of the integration by closing the integration contour.
Then we can rewrite Eq.(\ref{126}) after substitution  $l = i\tau$
in the form:
\begin{equation}
S^{-1}(0,R) = \frac{2}{(2\pi )^2\rho }\int_{0}^{\infty} d\tau e^{-\tau \rho}
\ \frac{\tau\ Im S^{mod}_{cor}(i\tau, X)}{\mid S^{mod}_{cor}(i\tau, X) \mid^2}.
\label{127}
\end{equation}
Applying the mean value theorem, the integral of
Eq.(\ref{127}) is estimated as:
\begin{equation}
S^{-1}(0,R) = C(T)\ \frac{{\rm e}^{-\mu TR}}{TR},
\label{128}
\end{equation}
\noindent
where $C(T)$ is a calculable quantity. The quantity $\mu$ is a
positive constant from the range of integration.  For our purposes here it is
sufficient to know that $\mu > 0$. The analytical evaluation of this
quantity shows its dependence on the chromoelectric mass.
Eq.(\ref{128}) is valid for
$R\neq 0$. The matrix element $S^{-1}(0,0)$ is infinite and represents the
quark self energy in the heavy quark potential given in Eq.(\ref{124}).
In what
follows we shall consider only the interaction energy of the quarks by
subtracting $S^{-1}(0,0)$ from all relations. We would like to point out the
importance of the Eq.(\ref{128}). All qualitative features of the following
discussion are based on the particular approximate form of the right-hand side
of Eq.(\ref{128}) and are independent on details of Eq.(\ref{100}).
Investigating the last equations we find the following picture.

For value of $X=\frac{1}{2}$  (which corresponds to the vanishing Polyakov
loop value $\langle L \rangle = 0$)
the heavy quark potential calculated from the
Eqs.(\ref{122}),(\ref{124}),(\ref{128}) possesses the desired $R$
dependence because for $R \rightarrow \infty$ the leading term is linear:
\begin{equation}
-\frac{1}{T}\{ V_{qq}^{interact}(R) \ - \ V_{qq}^{self\ energy} \} = -\mu RT +
const - \ln{RT} + O(R^{-2}).
\end{equation}
\noindent
In the deconfinement phase characterized by $\langle L \rangle \neq 0$
the effective potential possesses two
minima for $X \neq 1/2$ symmetric regarding to $X = 1/2$. The heavy quark
potential in this case is usually calculated by the
formula\cite{irb}:
\begin{equation}
-\frac{1}{T}V_{qq}(R) = \ln{\frac{\langle L(0),L(R)\rangle }{(\langle L(0)
\rangle)^2 }}.
\label{129}
\end{equation}
\noindent
Inserting Eqs.(\ref{124},\ref{128}) into the last equation we find for
$R \rightarrow \infty$ the result:
\begin{equation}
\frac{1}{T}V_{qq}(R) = tg^2(\pi X)\ \frac{g_r^2}{8}\ C(T)\ \frac
{{\rm e}^{-\mu TR}}{TR} \ + \ O(R^{-2}).
\end{equation}
\noindent
In the leading term of the last expression we recognize the Debye screening of
the colour charge potential in accordance with the lattice results\cite{irb}
as well as analytical calculations\cite{zin}. Thus we can see that
appearance of $X \neq n$ ($<A_0> \neq 0$)
leads to the Debye screened phase at high temperatures.

One of the most exciting consequences of $A_0$-condensate
is the spontaneous breaking of the colour charge symmetry and, consequently,
a generation of the baryonic number. If this phenomenon does realize in
nature it would be a good possibility both for verifying the general theory
described in this survey and for searching the new state of the strong
interacting matter.
It should be mentioned at once that the baryonic number is generated
not at all possible values of the condensate but only when the situation
described in the end of chapters 4, 5 is realized.
In $SU(2)$ gauge theory such a generation is not possible at all
because the appearance of the condensate leads to
the symmetric under charge conjugation effective action.
In $SU(3)$ theory the baryonic number is generated when both $<A_0^3>$ and
$<A_0^8>$ differ from zero.
The main statement is: {\bf In the presence of a non-vanishing Euclidean
$\langle A_0 \rangle$ field the generation of the nonzero baryonic number is
possible.} This statement was proved in two cases.

1) We studied the Hamiltonian formulation of the lattice QCD in strong
coupling limit \cite {bor2} where the partition function is of form (the
Kogut-Susskind fermions were used) (\ref{54}),(\ref{55}),
(\ref{103}):
\begin{eqnarray}
Z = \int\prod_{x}D\mu(A_0)\prod_{x,n}{\sum_{l}\exp(-\gamma C_{2}(l))
\Omega_{l}(A_{0}(x))\Omega_{l}^{\star}(A_{0}(x+n))}*  \nonumber  \\
\int\prod_{x,n}DU_{n}(x)Det[1+\exp(i\varphi^{\alpha}(x)) +
(Ta_{\sigma})^{-1}\sum_{n=d}^{-d}\eta_{n}(x)U_{n}(x)\delta_{x,y-n}]
\end {eqnarray}
\noindent
where $\gamma = g^{2}(2Ta_{\sigma})^{-1}$, $\Omega_{l}$ is character of the
l-th irreducible representation and the high-temperature expansion
in the Wilson loops was applied to calculate the fermionic determinant.
For each colour quantum number the corresponding charge density
generated by $\langle A_0 \rangle$ is given by:
\begin{eqnarray}
\rho_{\alpha} = \rho [(A_0)_{\alpha,\alpha}] =\lim_{\mu \rightarrow 0}(N)^{-1}
\frac{\partial\ln Z(g\varphi^{\alpha}\longrightarrow g\varphi^{\alpha}
\pm i\mu^{\alpha })}{\partial \mu^{\alpha }}.
\label{130}
\end{eqnarray}
\noindent
The chemical potential is introduced into the fermionic determinant. The
physical barionic density is then obtained as:
\begin{equation}
Q = \sum_{\alpha}\rho [(A_0)_{\alpha,  \alpha}],
\end{equation}
\noindent
 where $\alpha$ is the colour index.
After integrating over gauge fields we get the result:
\begin{equation}
\rho_{\alpha} = \langle \frac{Im\ W^{\alpha}(A_0)}{1+Re\ W(A_0)} \rangle +
(16T^{2}d)^{-1}\ \sum_{n} \langle \Delta^{\alpha} \rangle +O(T^{-4}), \
\Delta^{\alpha} \sim Im\ W^{\alpha}.
\end{equation}
\noindent
 From the last equation we can see that baryonic density is proportional to
the imaginary part of the Polyakov loop. It is easy to conclude that all the
higher orders of the $\beta$-expansion are also proportional to the
imaginary part of the fundamental characters. We point out now that
$\langle ImW \rangle$ will be different from zero if $\langle A_{0}^{3}
\rangle$ and $\langle A_{0}^{8} \rangle$ are non-vanishing.

2). The similar result was obtained in weak coupling regime
of the continuum QCD for free quarks being in the constant background
field $A_0$ \cite{pol2}.
In that case the baryonic number is expressed through different
representations of the imaginary part of the Polyakov loop:
\begin{eqnarray}
Q &=& i \sum_{\alpha}TV^{-1} \frac{\partial \ln Z}{\partial (A_0)_{\alpha,
\alpha}}  \nonumber  \\
&=& i \frac{2m^{2}T}{\pi^2}\sum_{n=1}^{\infty}\sum_{\alpha}(-1)^{n}
\frac{K_{2}(nm\beta)}{n} \sin(n\beta A_{0}^{\alpha}).
\label{131}
\end{eqnarray}
\noindent
Thus we can suppose that in the general case as well a non-vanishing
imaginary part of the Polyakov loop yields a non-zero baryonic density even
when baryonic chemical potential equals zero. It is very important
property of $\langle A_0 \rangle$-condensate which has to lead to
observable effects in the relativistic heavy ion collisions.

Now, we would like to draw an attention to the following problem
concerning the charge broken states. The free energy of $SU(3)$ gauge
theory has three global minima at the high temperatures.
On Fig.2 we have drawn them schematically for theory with dynamical quarks.
In the minimum "1" $\langle Tr\ ImW \rangle =0$ and, therefore, $Q =0$.
In other two minima the charge symmetry is broken,
$\langle Tr\ ImW \rangle \neq 0$ and, therefore, $Q \neq0$.
But we do not know in fact whether the C-symmetry broken states are
stable or metastable and, consequently, short-lived. There are some results
concerning this problem. One loop calculations ($\langle A_0 \rangle=0$)
with massless quarks have shown that the states where
$\langle Tr\ ImW \rangle =0$ are deeper \cite {dix}.
More precise 2-loop calculations with massless quarks (and with
$\langle A_0 \rangle \neq 0$) described earlier
lead to the same conclusions.
The lattice strong coupling evaluations
have confirmed	the such inference but calculations with massive quarks
\cite {bor} predict the existence of some values of parameters
$\gamma = \frac{\beta g^{2}}{2a},\ \beta,\ m\beta$ (considered as independent)
preferring the minima "2", "3" to be deeper than minimum "1" at
$\gamma \longrightarrow 0$. If it is the case we obtain the following
scenario.

In the chiral symmetric phase at high temperature $(m_{q} = 0)$ the main true
minimum of the free energy will be always real and other two will be
only metastable. If in the deconfinement phase the quarks
have a nonzero mass, the charge broken minima become deeper and the
baryonic number is nonzero. But at the chiral phase transition the system
will be back the real minimum again. Basically, in this case the
disappearance of the baryonic number and the transition of the system
to the state where $\langle Im\ TrW \rangle=0\
(\langle A^3_0 \rangle \neq 0,\ \langle A^8_0\rangle =0)$
is just the chiral phase transition. To verify this picture we
can look for the similar phenomena in $Z(3)$ spin system because most of
features of the $Z(3)$ spin system in d-dimension are similar to the $SU(3)$
ones at finite temperature\cite{yasve}.
We consider $Z(3)$ theory with external either positive or
negative magnetic field, whose partition function is of the form:
\begin{equation}
Z\ =\ \sum_{q_x =0}^{N-1}\ \exp \{\lambda \sum_{x,n}\
Re(e^{\frac{2\pi \imath}{N}q_x}\ \cdot \ e^{\frac{2\pi \imath}{N}q_{x+n}})\ +
\eta \sum_x \cos (\frac{2\pi}{N}q_x)\}
\label{132}
\end{equation}
\noindent
It is well known that\cite{unkn}:
\noindent
1) if $\eta >0$, the minimum is real, $\langle Im\ W\rangle=0$ and the phase
transition of the second order is possible at sufficiently small $\eta$;

\noindent
2) if $\eta<0$, the minima are complex ones, $\langle Im\ W\rangle \neq0$, the
phase transition of the Ising-type takes place to the state with broken
symmetry $q \rightarrow -q$.

What should be concluded from these points?
At least at the hopping-parameter
expansion of fermionic determinant the leading contribution including the
Polyakov loops has positive sign which definitely leads us to
the case $\eta >0$.
Of course, it is enough dangerous to do conclusions comparing QCD
with $Z(3)$ theory, nevertheless, it seems to us that the real
minimum will be deeper at any temperature in $SU(3)$ theory as well and,
hence, the states with nonzero baryonic number are, in fact, metastable.
In principle, we have not to except other possibility as well.
The situation can be much more complicated.
For example, the discussion in\cite{pol2} shows that
in QCD slightly above the deconfinement phase transition the domains of
different "vacua" (with different background $A_0$) may exist.
The scenario described above cannot be excluded, too, at least,
until we obtain the connection (or some dependence) between
$\langle A_0\rangle$  condensate and quark masses.

Another approach to resolve this problem has been conjectured in
\cite{faber}:
full QCD cannot be described by a grand canonical ensemble with
respect to triality or quark number. The vacuum of QCD with dynamical
fermions has triality zero and therefore degenerate $Z_3$ phases and
ordered-ordered phase transitions like pure gluonic QCD.
This means that full QCD does not
lead to metastable phases existing up to arbitrary high temperatures.

Certainly, it is very desirable to have more precise both analytical and
Monte-Carlo calculations concerning the metastable phases with massive and
massless quarks to clarify this very exciting problem.

The last question we would like to elaborate here is
an inducing of the Chern-Simons action in finite temperature gauge theories.
We adduce the sufficient conditions
of such a generation in the background $A_0$ field.

The Chern-Simons theories are very popular now owing to their
unusual and attractive properties.
The two areas are usually discussed to be relevant to
the Chern-Simons action: the superconductivity and 4-dimensional field theory
at finite temperature. There are some well-known examples
when the Chern-Simons action is induced by fermionic determinant
at high temperatures in four-dimensional theory.
Redlich\cite{red} has studied the model of two left-hand double fermions
connected with $SU(2)$ gauge field and imaginary chemical potential.
In \cite{ruth} QCD-like theory with axial $U(1)$ charges was studied.
In both cases the fermionic determinant generates at high temperature
the Chern-Simons action for the static modes of gauge field.
The coefficient in the front of the Chern-Simons term is
proportional to the introduced chemical potential.
In connection with it the following question is very interesting: is it
possible to induce the Chern-Simons term into finite-temperature QCD and what
is the mechanism of it?
At the first sight this question may seem a bit strange since there are
some problems which forbid appearance of the Chern-Simons term
in the QCD-action:

1) the Chern-Simons term is defined for $3$-dimensional theories;

2) $S_{c-s}$ violates the $CP$-symmetry.

This part of our survey is founded on the article \cite{bor3} where all these
problems are discussed in details.
In brief, the $S_{c-s}$ is induced for static
gauge modes which are described by appropriate	$3$-dimensional theory.
$CP$-symmetry is broken via certain regularization procedure on the lattice.
This breakdown vanishes identically in the naive continuum limit.
The Chern-Simons action is of the form
\begin {eqnarray}
S_{c-s} = 8\pi^2\kappa \int d^{3}x\ I(A),   \nonumber \\
I(A) = \frac{1}{8\pi^2}\varepsilon^{nmk}Sp(A_n\partial_{m}A_k +
\frac{2}{3}A_{n}A_{m}A_{k}).
\label{133}
\end {eqnarray}
Partition function has the standard form
\begin {equation}
Z = \int[DA_{\mu}]\exp \{-\int_{M}[dx]L_{G}-\Gamma_{eff}(A) \}
\end {equation}
where $\Gamma_{eff}(A)$ is the result of integration over quark fields.
We looked for action (\ref{133}) for smooth potentials of the static gauge
modes in $\Gamma_{eff}$ because we do not know how it is possible
to put down the action (\ref{133}) on the lattice in an arbitrary case.
We used the lattice regularization for the effective action and
studied both the Wilson and the Kogut-Susskind actions for
fermions with different boundary conditions for gauge fields.
Both the perturbative expansion and the expansion in the Wilson loops
of the determinant were applied to calculate $\Gamma_{eff}(A)$.
The summary of our results concerning $A_0$-condensate and
the Chern-Simons action is as follows:
(for detailed calculations see \cite{bor3}).

1. In perturbative expansion a coefficient at the Chern-Simons term
is expressed through free fermion propagator $G$,
\begin {equation}
\kappa = \frac{1}{6}\int_{-\frac{\pi}{a}}^{\frac{\pi}{a}}\frac{d^3p}{(2\pi)^3}
\varepsilon_{nmk}Sp[(G^{-1}\partial_{n}G)(G^{-1}\partial_{m}G)
(G^{-1}\partial_{k}G)].
\end {equation}

Analyzing this expression one can conclude that there is not any
perturbative relation between $A_0$-condensate and
the Chern-Simons action (though, exists
a possibility of $\kappa \neq 0$  for some values of the
Wilson parameter $R$; see below). Nevertheless, a nonperturbative connection
does exist.

2. Either for QCD on the torus due to certain periodic boundary conditions
for gauge fields
\begin {eqnarray}
M = (S^{1})^{d}:
A_{\mu}(x+l_{\nu}) = \Omega_{\nu}A_{\mu}(x)\Omega_{\nu}^{-1} +
\Omega_{\nu}\partial_{\mu}\Omega_{\nu}^{-1};  \nonumber  \\
\Omega_{\mu}(x+l_{\nu})\Omega_{\nu}(x) = \Omega_{\nu}(x+l_{\mu})\Omega_{\mu}(x)
\end {eqnarray}
\noindent
or if we take into account nonperturbative configurations
out of center of $SU(N)$,
the fermionic determinant will generate the $\theta$-term
$F_{\mu\nu}F_{\rho\lambda}\varepsilon_{\mu\nu\rho\lambda}$.
We suppose that the vortex configurations of gauge field,
existing on the lattice, can provide the same minimum of
the effective action as fields $A_{\mu} = 0$ \cite{stul}:
\begin {eqnarray}
Sp\ Z(\partial p)V(\partial p) = 1, \ Z(\partial p) \in Z(N),	  \\
V(\partial p) = V_{n}(x)V_{m}(x+n)V_{n}^{+}(x+n)V_{m}^{+}(x),  \nonumber  \\
V_{m}(x) \in SU(N)/Z(N).	 \nonumber
\end {eqnarray}

The $\theta$-term is reduced to the Chern-Simons action at high
temperatures and when $A_0$ condensate falls.
The coefficient at the action is $\kappa \approx \alpha n\beta \langle A_0
\rangle$ for the theory on the torus and $\kappa \approx \lambda \beta
\langle A_0 \rangle F_{nm}(@)$ for the second case.
Here, $n$ is the winding number of matrices $\Omega$
and $@$ is a vortex potential.

3. In theory with the $SU(3)$ gauge group and nonperturbative dielectric
vacuum $S_{c-s}$ is generated when expectation
value of the imaginary part of the Polyakov loop differs from zero.
Nontrivial dielectric vacuum appears in an effective three-dimensional
theory with the group having nontrivial center \cite{borpet}, \cite{ildgt}
(see also our discussion above).
In the last both cases the configurations from the center of $SU(N)$
are necessary for generation of $S_{c-s}$.
So it is very plausible that such a mechanism is impossible in the standard
continuum QCD where the variables from the center of $SU(N)$ are absent.
Besides, there is another necessary condition of the appearing
of $S_{c-s}$ here.
In any of these cases QCD has to belong to definite class of universality.
It means the following. Choosing the lattice fermionic action we had to solve
the problem of doubling the fermion degrees of freedom. We should use either
the
Wilson action \cite{wilson}
\begin {eqnarray}
S_W &=&
\frac{1}{2}a^{3}\sum_{x,\mu}[\overline{\Psi}(x)U_{\mu}(x)(R-\gamma_{\mu})
\Psi(x+a_{\mu}) +
\overline{\Psi}(x)U_{\mu}^{+}(x-a_{\mu})(R+\gamma_{\mu})\Psi(x-a_{\mu})]
\nonumber \\
 &+& a^{4}m\sum_{x}\overline{\Psi}(x)\Psi(x) - da^{3}\sum_{x}
\overline{\Psi}(x)R\Psi(x)
\end {eqnarray}
\noindent
or the Kogut-Susskind action \cite{kogut}
\begin {eqnarray}
S_{K-S} = \frac{1}{2}\sum_{x,n=-d}^{d} \eta_{n}(x)\overline{\Psi}(x)
U_{n}(x)\Psi(x+n) + ma\sum_{x}\overline{\Psi}(x)\Psi(x),     \\
\eta_{-n} = -\eta_{n}, \  U_{-n}(x)=U_{n}^{+}(x-n)  \nonumber
\end {eqnarray}
\noindent
which allow us to avoid undesirable degrees of freedom. The most general form
of the Wilson parameter R is \cite{12seil,kogut,zenk}:
\begin {equation}
R = s\exp(\imath\theta \gamma_{5})T, \ 0\prec s \leq 1, \ 0\leq \theta \leq
\pi,
T^{2} = 1,
\end {equation}
\noindent
where $T$ is some matrix from the fermionic space.
One of the $\eta$-symbols is \cite{symb}:
\begin {equation}
\eta_{n}(x) = (-1)^{x_{1}+x_{2}+...+x_{n-1}}\exp(i\pi[k(x+n)-k(x)])
\end {equation}
\noindent
where $k$ is any integer. At the level of free quantum theory
we cannot choose a unique form of these quantities.
More than that, it is just valid at any choice of $R$ and $\eta$:

- naive continuum limit is$(\theta,s,T,k)$-independent and chiral symmetry is
 restored;

- fermionic propagator and lattice Feynman's rules coincide in the continuum
limit with standard Feynman's rules;

- the property of the positivity is fulfilled.

Nevertheless, the theories with different $R$ can belong to different classes
of
universality since they can define different quantum continuum theories.
Examples of the such nonuniversality can be found in \cite{coste}.
In the only case if $\eta$-symbols or parameter $R$ satisfy some conditions,
for example,
\begin{eqnarray}
\theta \neq 0 , T = \gamma_{0} \nonumber   \\
\label{par}
\eta_{n}(x)\eta_{m}(x+n)\eta_{k}(x+n+m)\eta_{l}(x+n+m+k) = \varepsilon_{nmkl}
\\
n \neq m \neq k \neq l \nonumber
\end{eqnarray}
\noindent
the Chern-Simons action will be generated in QCD.
Fortunately this property does not depend on lattice regularization
\cite{coste}.
On the other hand it is very difficult task to prove that the $S_{c-s}$
is unique nonuniversal contribution to effective action in finite-temperature
QCD though it seems very plausible (see, for instance, \cite{coste}).
The parameters $R$ and $\eta$ as in (\ref{par}) violate the CP-symmetry.
As we could see above,
this violation disappears in the continuum theory both in
the naive limit and in the free quantum theory.
The appearance of $S_{c-s}$ is the only remnant
of this violation which survives in the continuum quantum limit.
There is only one theoretical reason for the choice of $R$ or $\eta$, the
property of the positivity \cite{zenk}.
But this property is satisfied for given $R$ and $\eta$.
So we may formulate the question: is it possible that QCD belongs to the
class of universality in which the Chern-Simons action exists? The answer
is left unclear so far. Certainly, the presence of $S_{c-s}$ has to lead
to interesting phenomena at high temperatures because $S_{c-s}$ can change a
statistics of a matter fields connected with it.
Thus, appearance of the Chern-Simons term in the QCD action at high
temperature is approved by the following complex of the circumstances:

a) appearance of $A_0$-condensate;

b) existing of nonperturbative vacuum which is formed by vortex potentials
from the center of the gauge group;

c) problem of universality.

To finish, we would like to notice that appearing of $S_{c-s}$ could lead to
solving of infrared problem as well, since the Chern-Simons action
generates magnetic mass just for the infrared dangerous static modes.

\section{Discussion and Summary. Unsolved problems}

In the present survey we have considered the mechanism of the spontaneous
breaking of the global gauge symmetry caused by the condensation of the gluon
field at high temperatures. We have discussed in details the most important
approximations for calculation of $A_0$ condensate both in the continuum
theory and within lattice gauge models.
We understand that it is almost impossible to give a
full review of all questions related to $A_0$ condensation
because breaking $SU(N)$ symmetry is certainly strong phenomenon
which has to reflect itself on many aspects of high temperature QCD.

As an essence of the above analysis we would like to stress once
more that in all considered approaches the gluon field condensation
at $T \neq 0$ has been determined and proved to be a gauge invariant
phenomenon. The nice and important fact that in all approaches we have
discussed here the gauge invariance of the condensate is manifested
by the different methods makes us sure that the condensation may
indeed be realized in the nature.
At the same time, a number of discrepancies in results
obtained by different methods (mainly as the role of quarks is
concerned) have been found. These points need to be investigated
separately in order to have more reliable results.

It is natural that in the every method of calculation there are own
problems which should be solved in the future. Generally
speaking, they can be divided in three categories dependently on
their importance and theoretical and practical significance. The main
problem now is to derive a conclusion on $A_{0}$ condensation with higher
loop contributions to be included. This is a complicated mathematical
task which should be investigated on the base of Nielsen's identities
in order to obtain a correct gauge invariant result. Then, if $<A_0> \neq 0$
will be determined finally, the problem of constructing the
consistent theory with $<A_0> \neq 0$ should be solved. This point should be
investigated in details because just here a number of theoretical
problems remain unsolved yet. Most essential of them is the
construction of partition function and calculation of various
observables with $<A_0> \neq 0$ to be taken into account.
 As we have seen, there are six minima
with the same depth in the $(x,y)$-plane and $Z(3)$ symmetry is broken.
This situation, but in the case when $Z(N)$ is preserved, has been
discussed in the paper \cite{pol2}.
 It was shown that an extension of standard
description of thermal equilibrium is needed when discrete
degenerate phases are to be included. However, how does it work in
the case of spontaneous breaking of $SU(N)$ has not been investigated yet.
In this line the investigations of infrared problem of gauge fields at
$T \neq 0$, the gluon magnetic mass, asymptotic behaviour of running
coupling constant are of great importance, too. Having these
questions answered one obtains a consistent theory with $<A_0> \neq 0$. After
that a phenomenology of finite temperature QCD will be more
transparent. And only this third of mentioned points may give a
possibility of a deriving of a final conclusion on gauge field condensate.
Monte-Carlo simulations and analytical evaluations on the
lattice making use some correlation inequalities could help in this problem
as well.

At last, the very interesting and exciting problem is to find a dependence
of $A_0$-condensate
on quark mass in order to clarify the possible connection between
$<A_0>$ and the order parameter of the chiral phase transition.
One may hope that such a connection really exists since the
$A_0$-condensate can drastically change the structure
of the quark propagator.
On this way we may also hope to find a solution of the problem of the
baryonic number generation.

O.A.B. thanks Institute of Physics of Slovak Academy of Sciences
for their kind hospitality where the main part of this review has been
prepared. The authors are much obliged to A.Nogova and S.Mashkevich
for careful reading of the manuscript.


\end{document}